\documentclass[twocolumn,twocolappendix]{aastex7}

\usepackage{multirow}
\usepackage{comment}
\usepackage{amsmath}

\newcommand{\triplet}{ICT-MJD59015}

\begin{document}

\title{The First Search for Optical Transient as a Counterpart of a Month-timescale IceCube Neutrino Multiplet Event}

\correspondingauthor{Seiji Toshikage}
\email{seiji.toshikage@astr.tohoku.ac.jp}

\author[0009-0002-5156-7819]{Seiji Toshikage}
\email{seiji.toshikage@astr.tohoku.ac.jp}
\affiliation{Astronomical Institute, Tohoku University, Sendai 980-8578, Japan}

\author[0000-0003-2579-7266]{Shigeo S. Kimura}
\email{shigeo@astr.tohoku.ac.jp}
\affiliation{Frontier Research Institute for Interdisciplinary Sciences, Tohoku University, Sendai 980-8578, Japan}
\affiliation{Astronomical Institute, Tohoku University, Sendai 980-8578, Japan}

\author[0000-0001-6857-1772]{Nobuhiro Shimizu}
\email{shimizu@hepburn.s.chiba-u.ac.jp}
\affiliation{International Center for Hadron Astrophysics, Chiba University, Chiba 263-8522, Japan}

\author[0000-0001-8253-6850]{Masaomi Tanaka}
\email{masaomi.tanaka@astr.tohoku.ac.jp}
\affiliation{Astronomical Institute, Tohoku University, Sendai 980-8578, Japan}
\affiliation{Division for the Establishment of Frontier Sciences, Organization for Advanced Studies, Tohoku University, Sendai 980-8577, Japan}

\author[0000-0003-2480-5105]{Shigeru Yoshida}
\email{syoshida@hepburn.s.chiba-u.ac.jp}
\affiliation{International Center for Hadron Astrophysics, Chiba University, Chiba 263-8522, Japan}

\author[0000-0002-0207-9010]{Wataru B. Iwakiri}
\email{iwakiri@chiba-u.jp}
\affiliation{International Center for Hadron Astrophysics, Chiba University, Chiba 263-8522, Japan}

\author[0000-0001-7449-4814]{Tomoki Morokuma}
\email{tomoki.morokuma@p.chibakoudai.jp}
\affiliation{Astronomy Research Center, Chiba Institute of Technology, 2-17-1 Tsudanuma, Narashino, Chiba 275-0016, Japan}
\begin{abstract}

Optical transients with timescale of months, such as supernovae (SNe) and tidal disruption events (TDEs), are candidates of high-energy neutrino sources. Multiple neutrino detections from the same direction within a month timescale provide a unique opportunity to identify such optical counterparts in the nearby Universe. In this work, we conduct archival search for the optical counterpart of an IceCube triplet event using the data of Zwicky Transient Facility. We develop a dedicated alert filtering system and validate the performance by following a blind analysis method. Applying this filtering system to the data after the detections of the IceCube triplet event, we find no transient candidates within the localization area. Assuming that the IceCube triplet event originates from an astrophysical source, we constrain parameters of optical transient, a peak luminosity and a decay timescale, using a simple signal model that is motivated by TDEs and superluminous SNe (SLSNe). Assuming the case with no time lag between neutrino detections and optical peak, almost entire parameter space of the known TDEs and SLSNe would be constrained. To give constraints on transients with a rapidly evolving light curve, quick follow-up observations for future  neutrino multiplet events are crucial.
\end{abstract}

\keywords{Neutrino astronomy (1100) --- Optical astronomy (1776) --- Transient sources (1851)}


\section{Introduction} \label{sec:intro}

High-energy astrophysical neutrinos are expected to be produced through the interactions of high-energy cosmic rays with surrounding matter and photons. Since neutrinos propagate directly toward us without interacting or being deflected, high-energy astrophysical neutrinos offer a unique opportunity to identify the sources of high-energy cosmic rays. The IceCube Neutrino Observatory, a cubic-kilometer-scale detector in Antarctica, has been detecting astrophysical high-energy neutrinos~\citep{ic2013PhRvL.111b1103A, ic2015PhRvL.115h1102A}. So far, multi-messenger observations have revealed two associations between high-energy neutrino signals and extragalactic persistent astronomical objects. One is an active galactic nucleus (AGN) with a jet pointing toward us, so-called blazar, TXS 0506+056, firstly identified associated with IceCube-170922A~\citep{IC2018txs}. The other is a nearby active galaxy NGC1068, identified as a neutrino point source by the excess at the position with 4.2$\sigma$ significance~\citep{IC2022ngc1068}. However, the dominant source of detected high-energy neutrinos is still unknown.

It is worth studying the contributions from the astrophysical transients. Indeed, several types of astrophysical transients have been suggested as the candidates for high-energy neutrino sources from theoretical point of view, such as gamma-ray bursts~\citep[GRBs,][]{1997WandB}, jetted supernovae~\citep[SNe,][]{2004razzaque}, SNe interacting with circumstellar material ~\citep[CSM interacting SNe,][]{2011murase}, and tidal disruption events ~\citep[TDEs,][]{wang2011PhRvD..84h1301W}.

To study the association between high-energy neutrinos and these astrophysical transients, several works performed optical follow-up observations searching for the counterpart of high-energy neutrino detections such as Dark Energy Survey~\citep[DES,][]{2019morgan}, All-Sky Automated Survey for Supernovae~\citep[ASAS-SN,][]{2022necker} and,  Zwicky Transient Facility~\citep[ZTF,][]{2023stein}. As a result, associations of IceCube neutrino events with TDEs~\citep{stein2021, reusch2022, 2023jiang, 2024yuan} and a marginal connection with a CSM interacting SN ~\citep[Type Ibn,][]{stein2023} have been reported. 

While these previous works have provided valuable insights about the connection between transients and high-energy neutrinos, there are still several shortcomings. 
These studies relied on the IceCube neutrino alerts for single neutrino detection events (``singlet") whose redshift distributions extend to a distant Universe. For example, assuming singlet from a transient with neutrino emission energy $\varepsilon_{\rm \nu}=3\times10^{49}~{\rm erg}$, the redshift distribution is expected to extend up to $z \geq 2$ \citep{2022yoshidaApJ...937..108Y}. So far, optical transient counterpart searches for singlet events have never covered such a high redshift range. Moreover, such deep follow-up observations suffer from a large number of backgrounds from unrelated objects due to the large survey volume. 

To overcome this issue, we focus on ``multiplet", multiple neutrino detections from the same direction within a certain time window \citep{2017icmultiplet, 2022yoshidaApJ...937..108Y}. Multiplet sources emit neutrinos in a short timescale and thus, the astrophysical sources are likely to be transient phenomena. 
Moreover, the sources are expected to be located at relatively low redshift ($z \lesssim 0.3$, see Section \ref{sec:ana} for more details). Therefore, transient surveys focusing on relatively nearby Universe, such as the ZTF, can cover the expected redshift range of multiplet neutrino sources.

In this study, we perform archival search for an optical counterpart of an IceCube ``triplet" event (hereafter referred to as \triplet)~identified by \citet{2025shimizu}. \citet{2025shimizu} searches multiple neutrino detections with the time window of $T_{\rm w}=30~{\rm days}$ focusing on a relatively long time scale transients such as SNe and TDEs. For the optical counterpart search, we use ZTF archival data. By following a blind analysis approach, we determine the signal detection criteria without analyzing the data in the direction of the target IceCube triplet event. After establishing these criteria, we analyze the data in the direction to search for the optical counterpart. Based on the results, we perform a statistical analysis to give constraints to the properties of neutrino-emitting transients.

This paper is structured as follows. In Section~\ref{sec:data}, we give the details of the IceCube triplet event \triplet~and the ZTF data used for optical counterpart search. We describe our analysis strategy in Section~\ref{sec:ana}. We present the results from our analysis in Section~\ref{sec:result}. In Section \ref{sec:discussion}, we discuss  implications from our results. Finally, we give summary of this work in Section~\ref{sec:summary}.

\section{Data} \label{sec:data}

\subsection{IceCube Data} \label{sec:data:ic}

The IceCube collaboration recently reported a search for month-long 
doublets and triplets 
in 11.4 years of data~\citep{2025shimizu}. 
Track-type neutrino events, which are primarily initiated by muon neutrinos, 
are selected in Northern Sky high-purity neutrino sample~\citep{gfu_2016}. 
The main background originates from atmospheric neutrinos, and 0.3\% of events in the 
dataset are from astrophysical origin. For the selected neutrino events, 
significant two (doublet) or three (triplet) 
event pairs in a time window of $T_{\rm w}=30~{\rm days}$ were selected 
based on their energies and consistency of their directions. 
The pairing scheme was optimized for 
small number of neutrino detections, and the sensitivity to dim sources was 
improved by up to a factor of two compared to 
previous methods~\citep{Kintscher2020Rapid,gfu_2016}. 
The contribution of dominant atmospheric neutrinos is 
suppressed as their energies increse, 
and the signal becomes significant for $E_\nu \gtrsim 50$~TeV.

As a whole dataset, the observed multiplets were consistent 
with background-only hypothesis, and the 
null observation provided a constraint on $\varepsilon_\nu$ and 
burst rates of transients. Nevertheless, the two most significant 
multiplets had false alarm rates (FARs) 
smaller than the inverse of the 
total live-time ($1/11.4~\mathrm{yr}^{-1}$), 
implying a possible excess.
The directional uncertainty of the source candidates of the 
two multiplets were presented in the paper, and 
showed excellent localizations within a radius of $\sim 0.3^{\circ}$ at 90\% containment. 
Of these two multiplets, 
ZTF data (explained later in Sec.~\ref{sec:data:ztf}) covered the 
localization area of the second most 
significant triplet, \triplet, after the last neutrino detection. 
From Figure~4 of~\citealt{2025shimizu}, 
when a neutrino emission energy per transient 
source is $\varepsilon_\nu=10^{52}~\mathrm{erg}$ and their 
local volumetric event rate is $R_0 = 10^{-8}~\mathrm{Mpc}^{-3}\,\mathrm{yr}^{-1}$ (these choices 
are inferred with the cosmic neutrino background flux measured
by IceCube, \citealt{Abbasi_2022}), 
a signalness---a probability that \triplet~was from astrophysical origin---can be inferred to be approximately 60\%. Here, signalness is defined as the fraction of the multiplet originating from the astrophysical transient source to the total number of multiplets with identical test statistics as the triplet in question. The test statistic is a metric that is employed to evaluate the astrophysical hypothesis in comparison to the atmospheric neutrino hypothesis. This evaluation is made under the assumption that the observed cosmic neutrinos are generated by transient sources (see Appendix \ref{ap-signalness} for the detail). Although it was pointed out that there was a 4FGL~Fermi 
source in the uncertainty region of the triplet~\citep{Abdollahi_2022_fermicatalog, ballet2023fermi_DR4}, the association was likely an accidental coincidence based on the inconsistency 
between IceCube sensitivity and the distance to the source ($z=0.46$). 
Table~\ref{table:triplet} summarizes the information of \triplet\,~\citep{2025shimizu}.

\begin{table}[t]
\begin{center}
\caption{Information of \triplet \label{table:triplet}} 
\centering
\scalebox{0.9}[0.9]{
\hspace{-2cm}\begin{tabular}{cccc}  \hline \hline
Type & \multicolumn{3}{c}{Triplet} \\
FAR & \multicolumn{3}{c}{$0.078~\mathrm{yr}^{-1}$} \\
Direction& \multicolumn{3}{c}{ (RA, DEC) $=(0.58^{\circ}$, $-0.35^{\circ})$}\\Energy$^\dagger$ & \multicolumn{3}{c}{$\log_{10} E_{1}=5.47, \log_{10} E_{2}=4.31, \log_{10} E_{3}=3.62$} \\
\hline
Time & $t_{\nu,1}$ & $t_{\nu,2}$ & $t_{\nu,3}$ \\
\hline
MJD& 59011.22 & 59015.46 & 59027.66 \\
UT& 2020-06-11.22 & 2020-06-15.46 & 2020-06-27.66 \\
\hline \hline
\end{tabular}
}
\end{center}
{\vspace{-2mm} \footnotesize $\dagger$ Energy is in GeV.}
\end{table}

\begin{figure}[t!]
\epsscale{1.1}
\plotone{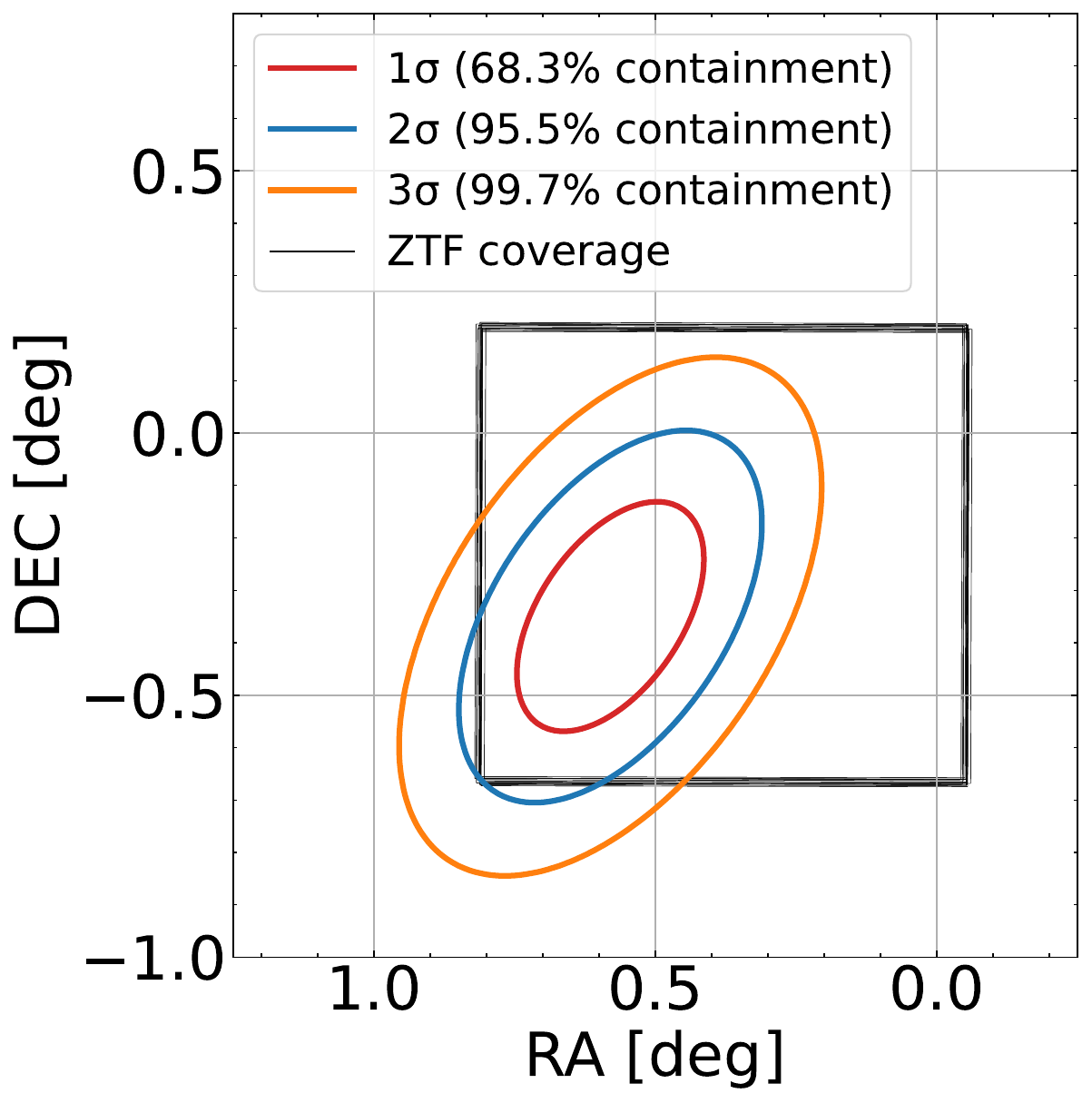}
\caption{The direction of IceCube triplet event \triplet~and ZTF coverage for the field = 396, ccdid = D15, and qid = 3. Here qid is the ID of the quadrant for each read-out area.}
\label{fig:localization}
\end{figure}

\subsection{ZTF data} \label{sec:data:ztf}

We perform archival search for optical transient associated with \triplet \ by using ZTF data \citep{bellm2019PASP..131a8002B}.
ZTF performs an optical transient survey using a wide-field camera with 47 deg$^2$ field of view mounted on the Palomar 48-inch Schmidt telescope.
The camera equips 16 of 6k $\times$ 6k CCDs. 
The imaging data for each CCD are read out into 4 of 3k $\times$ 3k image. 
The survey is mainly performed with optical $g$- and $r$-band filters (mean wavelengths of 4800 \AA \ and 6500 \AA, respectively).
The cadence of the public data is typically about 3 days.

The ZTF survey started to cover the localization area of the triplet event since July 6, 2020 (25 days after the detection of the first neutrino event).
The black square in Figure \ref{fig:localization} shows the coverage of the ZTF data around \triplet: a quadrant of 1 CCD chip for the field = 396, ccdid = 15, and qid = 3, covering 0.83 $\times$ 0.83 $\mathrm{deg}^2$. When we perform signal model test (Section \ref{sec:ana:test}), we take into account an effect of the fractional coverage of the neutrino error region.
After July 6, 2020, 58 and 57 pointing observations have been performed by the end of 2020 in $g$- and $r$-bands, respectively.
The median 5$\sigma$ limiting magnitudes are 21.3 mag and 21.5 mag in $g$- and $r$-bands, respectively.
Since the first data taken after the IceCube triplet event, on July 6, 2020, are rather shallow (with a 5$\sigma$ limiting magnitude of $< 20$ mag), we use the data after the second date, July 14, 2020.

By the ZTF data reduction pipeline \citep{2019PASP..131a8003M}, transient candidates are selected by performing image subtraction using the deep reference image.
Then, sources with $>5 \sigma$ detection in the subtracted image are publicly distributed through the alert stream\footnote{https://ztf.uw.edu/alerts/public/} \citep{patterson2019PASP..131a8001P}. 
Note that an ``alert" is issued for each detection, and a certain ``object" is associated with multiple ``alerts".

As the alerts in this stage are dominated by bogus detections mainly due to residual by imperfect image subtraction, pixels around bright stars, and cosmic ray hits, ZTF provides a recommended set of quality cut about the shape of the detected sources (see Section \ref{sec:ana} for more details).
Then, a further cut is applied to remove solar system (moving) objects by cross-matching with known asteroids and by imposing multiple, positive detections at the same position. Finally, to remove Galactic variable stars, the nature of the closest objects is evaluated based on the “star-galaxy separation score” ({\tt sgscore} $= 1$ means more star-like while {\tt sgscore} $= 0$ means more galaxy-like, \citealt{tachibana2018PASP..130l8001T}). 
An object detected in Pan-STARRS1 (PS1) image within 1.5 arcsec from the transient candidate is assigned as a host object.
If the host object is a galaxy-like object with {\tt sgscore} $<0.5$, such a candidate passes the selection.

The recommended filtering criteria successfully work for the real-time extragalactic transient search.
However, these cannot be readily used for our neutrino follow-up for the following reasons.
First, the recommended conditions above apply 
a strict condition of {\tt sgscore}
to greatly 
reduce the number of stellar-like host objects. 
However, this condition may filter out some real transients in galaxy-like host objects.
Second, the overall true positive rate (TPR) or passing rate of real transients (i.e., true signal) of these filtering conditions is not quantified.
Moreover,
the number of unrelated objects passing these filters, i.e.,  a background rate in the neutrino follow-up search, has never been quantified.
These prevent an evaluation of the statistical significance of the detection (or non-detection) of optical transients associated with neutrino events.
Therefore, we develop a new alert filtering scheme dedicated for neutrino follow-up observations and carefully measure the TPR of true transients and background rate, as described in Section \ref{sec:ana}.

\section{Analysis} \label{sec:ana}

\subsection{Analysis strategy} \label{sec:ana:strategy}

\label{sec:Ana:strategy}

In this work, we follow a blind analysis strategy. This means that we do not analyze the data in the localization area of \triplet\ (Figure~\ref{fig:localization}) until we fully determine the criteria of our alert filtering scheme. 

We define signal as the optical transient associated with the neutrino triplet event \triplet. It is expected to be an extragalactic transient with a timescale of months at a relatively low redshift (see Section~\ref{sec:ana:model}). 
The background of our analysis is divided into two classes: bogus detections and astrophysical background events. Bogus detections consist of artifacts arising from image subtraction. Astrophysical background events are unrelated optical transients such as variable stars, AGNs, unrelated SNe and TDEs in the direction of \triplet. 

To distinguish the signal from these backgrounds, we develop our alert filtering system. 
For the statistical analysis, we must carefully measure the TPR of our filters
and the expected number of background objects retained in the final sample.
For the estimation of the TPR of signal, we adopt a hybrid approach combining an analytical source model and ``data-driven" method. Since the characteristics of optical transients associated with the triplet event \triplet, such as luminosity and timescale are not known in advance, we implement the light curve model by employing an analytical function that possesses specific parameters designed to describe the characteristics of the source 
(see Section~\ref{sec:ana:model}). 
Nevertheless, the TPR is contingent upon the actual image quality, which is subject to the observing conditions.
Thus, it is necessary to use the actual imaging data (i.e., ``data-driven" method) 
to reliably estimate the TPR. 
To this end, we construct control data samples, 
which consists of ZTF imaging data obtained from sky regions 
that do not lie within the \triplet~localization region
(see Section~\ref{sec:ana:bg} and \ref{sec:ana:selection}). 
The control data is then utilized to empirically estimate 
the number of alerts or objects that pass the series of signal selections, 
which is then turned into the TPR estimate following the data-driven approach
(see Section~\ref{sec:ana:multi}). The estimation of the number of backgrounds is made entirely data-driven, with the control data set serving as the primary source of information.
The estimated TPR and background penetration rate 
are used as inputs in a statistical analysis to constrain the characteristics of the neutrino transient source (see Section \ref{sec:ana:test}). 
The TPR and background rate for each stage of the filtering levels are summarized in Table~\ref{tab:summary}.

\subsection{Model of signal - optical emission from $\nu$ source}
\label{sec:ana:model}

In this section, we describe our signal model, i.e., the light curve model of the optical emission from the neutrino source. We here do not consider a specific neutrino emission model from optical transient. We provide a simple optical lightcurve model for the neutrino source to constrain the parameter space of optical signals. Implications on specific scenarios are discussed in Section \ref{sec:discussion}.

We assume that the source has an exponentially decaying lightcurve with a sharp rise at the source frame:
\begin{equation}
    L_\nu(t) = L_{\rm pk}\frac{B_\nu (T)}{\sigma T^4 /\pi}\exp\left(-\frac{|t-t_{\rm pk}|}{t_{\rm decay}}\right)\Theta(t-t_{\rm pk}),
\end{equation}
where $L_\nu$ is the differential photon luminosity (per frequency), $L_{\rm pk}$ is a normalization factor, $t_{\rm pk}$ is the peak time of the lightcurve, $t_{\rm decay}$ is the decay time, $B_\nu (T)$ is the Planck function, $T$ is the temperature of the thermal photons, and $\Theta(x)$ is the Heaviside step function. Since we analyze the data after the neutrino detections, the rise phase of model light curve does not affect our analysis (see Section~{\ref{sec:ana:multi}}). 

In this study, we choose $t_{\rm decay}$ and $L_{\rm pk}$ as primary parameters. 
For simplicity, we adopt $T=10^4$ K and $2\times 10^4$ K for superluminous SNe (SLSNe) and TDEs respectively, based on the typical blackbody temperatures of known SLSNe and TDEs \citep{vanvelzen2021ApJ...908....4V, yao2023ApJ...955L...6Y, chen2023ApJ...943...41C}. For SLSNe, we also tested other choice of temperatures and confirmed that the results remain largely unchanged as long as $T\gtrsim5000~{\rm K}$.
We examine two values of the peak time of the optical lightcurve: $\Delta t=0$ and 30 days, where $\Delta t=t_{\nu, 1}-t_{\rm pk}$.

We model the redshift distribution of the source as follows. First, we give the local volumetric event rate of a neutrino source class, $R_0$. The current IceCube sensitivity for neutrino multiplet detection
allows us to examine the objects with $R_0\lesssim 10^{-7}\rm~Mpc^{-3}~yr^{-1}$ \citep{2022yoshidaApJ...937..108Y}, which is comparable to the events rates of TDEs \citep[e.g.,][]{yao2023ApJ...955L...6Y} and SLSNe \citep[e.g.,][]{moriya2019ApJS..241...16M}.
More frequent transients, such as interaction-powered SNe, 
are unlikely to yield detectable neutrino multiplet events
with the current neutrino detector.

Suppose that a single source class provides the dominant contribution of the cosmic neutrino background intensity measured by IceCube. Considering neutrino emission from standard candles (i.e., the sources with the same luminosity contribute the cosmic background at any redshift), we can estimate the mean neutrino fluence from a source by 
\begin{equation}
    \varepsilon_\nu \sim \frac{12\pi  H_0\int E_\nu\Phi_\nu dE_\nu}{c \xi_z R_0 }, 
\end{equation}
where $\Phi_\nu$ 
is the cosmic muon neutrino background intensity \citep{icecube2022ApJ...928...50A},  $\xi_z
=H_0^{-1}\int dz (1+z)^{1-\gamma} (dt/dz)\psi(z)$ 
is the correction factor by the cosmological source density evolution, $\gamma$ is the neutrino spectral index,
$\psi(z)$ is the redshift evolution of the source, 
$H_0$ is the Hubble constant, and $c$ is the speed of light. 
Given the redshift evolutions of TDEs  \citep{sun2015ApJ...812...33S, 2025necker} and SLSNe \citep[e.g.,][]{2022yoshidaApJ...937..108Y} with the observed neutrino spectral index $\gamma\simeq 2.3$, we obtain $\xi_z \sim 0.3$ and 2.5, respectively. 
We consider two cases for TDEs given the uncertainty of the event rate, and one case for SLSNe. The mean neutrino energy fluence is estimated to be $\varepsilon_\nu\simeq 1.5\times10^{52}$ erg for TDE with $R_0=6\times10^{-8}\rm~Mpc^{-3}~yr^{-1}$ (TDE-L; $R_0$ is given by the break in the luminosity function by \citealt{yao2023ApJ...955L...6Y}), $\varepsilon_\nu\simeq 3\times10^{51}$ erg for TDE with $R_0=3\times10^{-7}\rm~Mpc^{-3}~yr^{-1}$ (TDE-F; the case where fainter TDE emit neutrinos),  and $\varepsilon_\nu \sim 1.5\times10^{51}$ erg for SLSNe with $R_0=1\times10^{-7}\rm~Mpc^{-3}~yr^{-1}$ (SLSN; \citealt{moriya2019ApJS..241...16M,curtin2019ApJS..241...17C}).

\begin{figure}[t!]
\epsscale{1.1}
\plotone{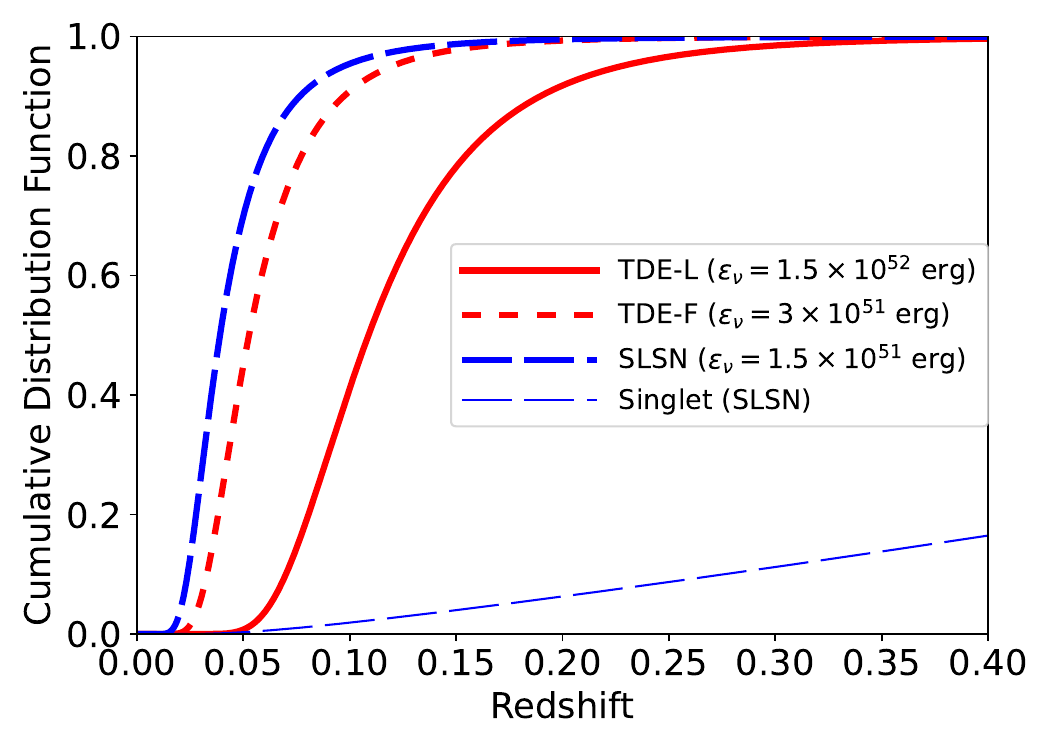}
\caption{Cumulative probability density of redshift distribution for the IC neutrino triplet sources depending on the total neutrino energy $\varepsilon_\nu$.
}
\label{fig:zpdf}
\end{figure}

Using these values of $\varepsilon_\nu$ with the source redshift evolutions, we estimate the neutrino flux from each source and compute the Poisson probability of detecting it as a triplet event. By integrating over redshift, we calculate the number of triplet sources and the probability distribution functions of the sources of triplet events, $P_{\rm dist}(z)$ as shown in Figure \ref{fig:zpdf} (see Eqs. (8) and (16) in \citealt{2022yoshidaApJ...937..108Y}). 
Independently, we can evaluate the detection horizon of the optical emission from the neutrino source, $z_{\rm lim}$, given the values of $L_{\rm pk}$, $t_{\rm decay}$, $t_{\rm pk}$, $T$, and the sensitivity of ZTF. 
For luminous transients of $L_{\rm pk}\sim10^{44}\rm~erg~s^{-1}$ that we are targeting in this study, $z_{\rm lim}\sim0.4$ is achieved with ZTF. 
In this case, the probability that the source is located within $z_{\rm lim}$, i.e., $f(z<z_{\rm lim}) = \int^{z_{\rm lim}} P_{\rm dist}(z)dz$, is estimated to be $>0.999$ for triplet events while it is $\sim0.2$ for singlet events.
In the case of non-detection of any astrophysical transient within $z_{\rm lim}$, this probability corresponds to the significance level to exclude the adopted parameters of the transient.
The high $f(z<z_{\rm lim})$ demonstrates the advantage of follow-up of triplet events: thanks to the preference toward a lower redshift, a stronger constraint can be obtained as compared with singlet events.

\subsection{Background modeling} \label{sec:ana:bg}

The empirical modeling of the background penetration is based on the control data samples constituted by the ZTF wide-field data. To this end, three samples are created for the purpose of investigating different selection processes. The Wide Field (WF) control data sample comprises observation data collected during a single night on July 30, 2020, in the $g$-band, encompassing a total of 126,645 ZTF alerts.
The sky coverage is about $6500~\rm{deg^2}$ (see also Appendix~\ref{ap1}),
excluding the localization area of \triplet. 
The remaining two samples contain data collected during a series of observation epochs. These data are used to calculate the expected numbers of background and TPRs
resulting from examining multiple observations. 
The Multiple Epoch control data sample (hereafter referred to as ME1) consists of data from 11 different epochs spanning from July 14 to July 31 of 2020,
covering $300~\rm{deg^2}$. The ME2 sample is constituted by data from 6 epochs 
ranging from July 14 to July 23 with $750~\rm{deg^2}$ coverage. We use all the area observed in these epochs except for Galactic plane (Galactic latitude $|b|<10~\rm{deg}$), whose characteristic are very different from those of the target area of the triplet event. See Section~\ref{sec:ana:multi} for more details about this multi-epoch analysis.


These control samples contain the noises caused by bogus detections and astrophysical transients unrelated to neutrino emissions, which are two representative classes of the background events in the search for the neutrino counterpart. The reduction rate of these backgrounds is obtained by applying each of the event selection criteria to the control samples. The corrections for the sky patch difference between the control data samples and the region of \triplet~localization are subsequently employed to calculate the expected number of backgrounds that remain after all the event selections.


\begin{table*}[htb]
  \centering
    \caption{Summary of the performance of our filtering system. 
    \label{tab:tpr-nbg}
    }
  \begin{tabular}{cccrrr} \hline \hline
    \multicolumn{6}{c}{\triplet}  \\
    \hline
    Selection & & TPR & \multicolumn{3}{c}{Number of background} \\
    & & & WF & WF\phantom{0.12} & ME2\phantom{ 0.12} \\
    & & &  & $(\rm{deg^{-2}})$\phantom{0.1}& $(\rm{deg^{-2}})$\phantom{0.1} \\
    \hline 
    All 5$\sigma$ detections	&  	&  	$ - $		& 126,645 & $19.5\pm{0.1}$\phantom{23} & $-$\phantom{ 0.123}\\
    \hline 
    \multirow{ 2}{*}{\shortstack{level 1 selection \\ (removing bogus detections})}	 & $g$-band & $0.628\pm{0.010}$ 	&  \multirow{ 2}{*}{33,339} & \multirow{ 2}{*}{$5.13\pm{0.03}$\phantom{3}} & $-$\phantom{ 0.123}\\
     & $r$-band & $0.578\pm{0.009}$ 	& &  	& $-$\phantom{ 0.123}\\
    \hline
    \multirow{2}{*}{\shortstack{level 2 selection \\ (
    removing moving objects})} & $g$-band & $0.948\pm{0.007}$ & \multirow{ 2}{*}{11,622} & \multirow{ 2}{*}{$1.79\pm{0.02}$\phantom{3}}  & $-$\phantom{ 0.123}\\
    &  $r$-band & $0.995\pm{0.003}$ 	&  & 	& $-$\phantom{ 0.123}\\
    \hline
   \multirow{ 4}{*}{\shortstack{level 3 selection \\ (removing known variable sources)}} &  {\tt sgscore} & $0.9950\pm{0.0004}$ &  1,809 & $0.279\pm{0.007}$& $0.241\pm{0.017}$\\
   &  Gaia variable & $0.9987\pm{0.0004}$& 449 & $0.069\pm{0.003}$	& $0.111\pm{0.012}$\\
   &  Gaia proper motion & $0.9985\pm{0.0004}$& 416  & $0.064\pm{0.003}$& 
   $0.102\pm{0.011}$\\
   &  AGN & ${0.9973\pm{0.0005}}$& 397 & $0.061\pm{0.003}$& $0.094\pm{0.011}$\\
       \hline
          \multirow{ 2}{*}{\shortstack{level 4 selection \\ (removing variable sources using light curves)}} &  \multirow{ 2}{*}{Duration ($\Delta t_{\rm neg}$)} &  \multirow{ 2}{*}{${0.9977\pm{0.0016}}$} 	&  \multirow{ 2}{*}{138} &  \multirow{ 2}{*}{$0.021\pm{0.002}$}&  \multirow{ 2}{*}{$0.084\pm{0.010}$}\\
          \\
          \hline
      \end{tabular}
    \label{tab:summary}
    \begin{minipage}{18cm}
    \vspace{0.1cm}
\small  {\bf Notes}: For the level 1 and level 2 selections, we show the TPR for detections in typical magnitude range (18.5-19.0 mag). The $1 \sigma$ uncertainties reported for TPRs are calculated based on binomial distribution. For the number of background objects, we show the cumulative number of alerts for the detection-wise criteria (level 1 and level 2) and cumulative number of objects for the object-wise criteria (level 3 and level 4) with the $1\sigma$ statistical uncertainties  ($\pm{\sqrt{N}}$ of the original numbers of background objects).
\end{minipage}
\end{table*}

\begin{figure*}[t!]
\epsscale{1.1}
\plotone{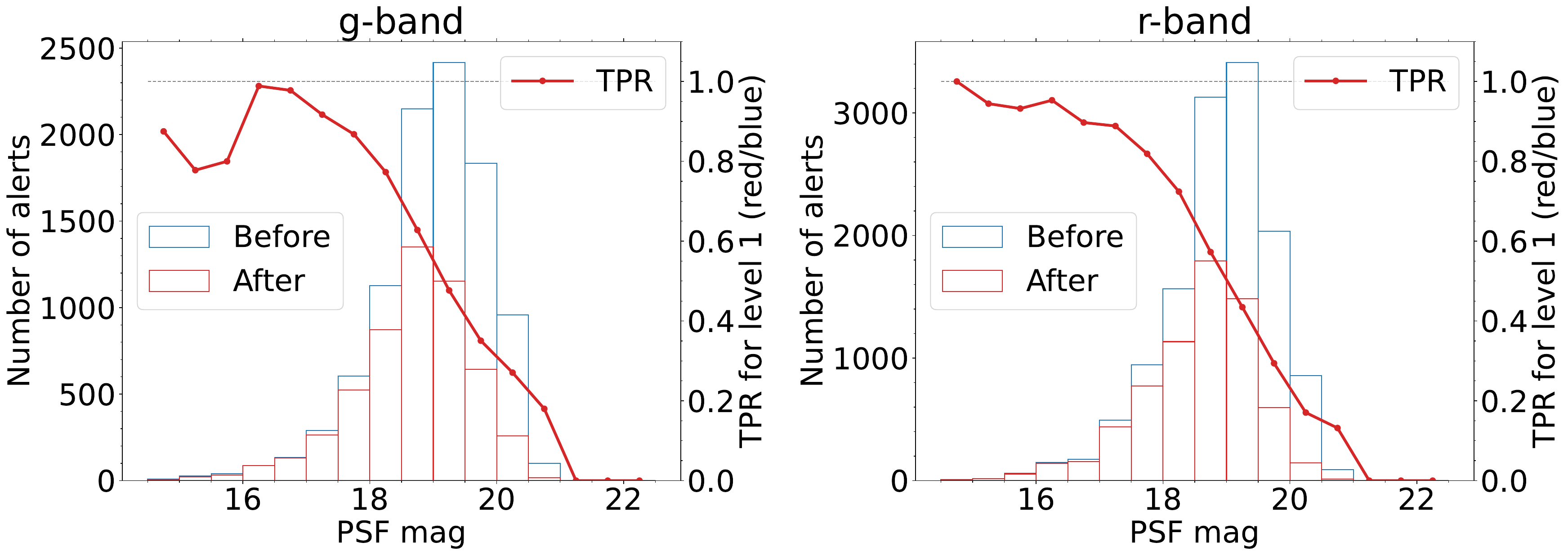}
\caption{True positive rate of level 1 selection estimated with alerts associated with transients reported in BTS. Histograms represent the number of alerts in each magnitude bin before/after level 1 selection (blue/red). The red solid line represents the TPR in each magnitude bin.}
\label{fig:tpr-qc}
\end{figure*}

\begin{figure*}[t!]
\epsscale{1.1}
\plotone{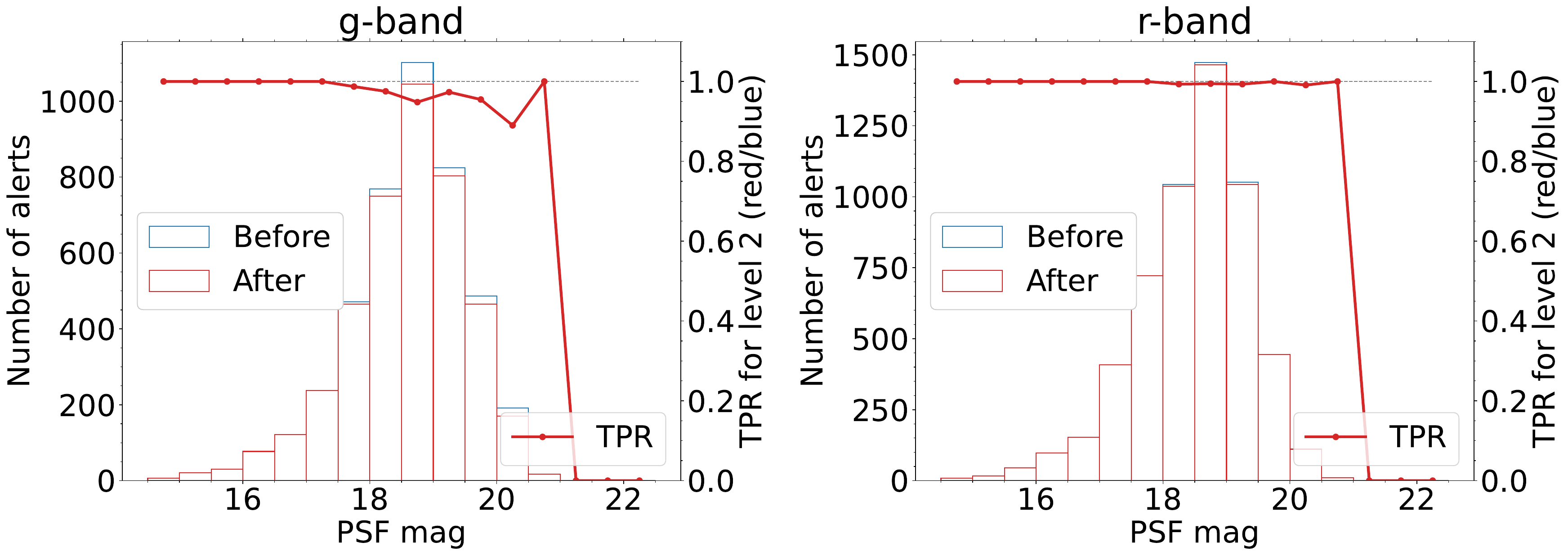}
\caption{True positive rate of level 2 selection estimated with alerts associated with SNe reported in BTS. Histograms represent the number of alerts in each magnitude bin before/after level 2 selection (blue/red). The red solid line represents the true positive rate in each magnitude bin.}
\label{fig:tpr-ec1}
\end{figure*}

\subsection{Event Selections}
\label{sec:ana:selection}

\subsubsection{level 1 selection}
\label{sec:ana:QC}

The level 1 selection is detection-wise cut to remove bogus detections. 
For this selection, we impose the following criteria on the data quality and shape of detected sources as suggested by the ZTF team\footnote{https://zwickytransientfacility.github.io/ztf-avro-alert/filtering.html}:
the number of bad pixel ({\tt nbad} $= 0$), the full width at half maximum (${\tt fwhm}<5~{\rm pixels}$), elongation (${\tt elong}\le 1.2$), and the difference between aperture magnitude and point spread function (PSF)-fitted magnitude (abs({\tt magdiff}) $< 0.1$). 
Additionally, the ZTF team developed a machine learning Real/Bogus (RB) classifier \citep{2019mahabal} which estimates the RB score for individual detection ranging from 0 (bogus-like) to 1 (real-like). We adopt the threshold for RB score suggested by ZTF (${\tt rb} \geq 0.65$).

We estimate the TPR of this level 1 selection using the alerts of real transients. We use real transients reported by the Bright Transient Survey~\citep[BTS,][]{fremling2020ApJ...895...32F, perley2020ApJ...904...35P}, which is a magnitude-limited ($m < 19$ mag in either the $g$- or $r$-band in peak) survey of extragalactic transients. 
For the transients reported by BTS, we collected 22,733 associated alerts from May 2020 to August 2020. The alerts associated with BTS sample cover a wide magnitude range down to 21 mag.
The TPR is estimated by the number of alerts passing the level 1 selection divided by the number of all the alerts associated with the transients reported by BTS. Figure~\ref{fig:tpr-qc} shows TPR along with PSF magnitude of each alert. 
For a typical magnitude range ($18.5\leq \rm{PSF mag} \leq 19.0$), the TPR of the level 1 selection is $0.628\pm{0.010}/0.578\pm{0.009}$ in $g$/$r$-band  ($1\sigma$ statistical uncertainties are calculated based on binomial distribution). 

We also estimate the number of background alerts penetrating the level 1 selection. 
Applying the selection to WF control data sample, 126,645 alerts are reduced to 33,339 alerts by the level 1 selection.
This corresponds to a background rate of $5.13\pm0.03$ alerts~$\rm deg^{-2}$ (the error is based on the statistical uncertainties of $\pm{\sqrt{N}}$ for the original numbers of background objects, assuming Poisson statistics).


\subsubsection{level 2 selection}
\label{sec:ana:EC1}

The level 2 selection is designed to remove moving objects in the Solar system while retaining candidates for new  transients. 
The criteria applied in the standard ZTF alert flow 
are employed in this case as well: 
(1) no known solar system objects within 5 arcsec, (2) at least two $>5 \sigma$ detections separated by $> 30$ min, and (3) the detection with a positive flux in the subtracted image. 

We estimate the TPR of the level 2 selection in the same way as the level 1 selection. 
However, the TPR against the criterion (2) is contingent upon the phase of the light curve. Consequently, we consider the contribution from this criterion at the subsequent stage when performing the statistical test on the neutrino transient source model (see Section~\ref{sec:ana:test}) by imposing the criterion of multiple detections. The TPR that corresponds to the level 2 selection criteria, except criterion (2), is determined using the same data-driven approach employed for the TPR calculation in the level 1 selection. We apply the level 2 cuts to the alerts of real BTS transients to probe the TPR.

Figure~\ref{fig:tpr-ec1} shows TPR as a function of PSF magnitude in each band. For a typical magnitude range ($18.5\leq \rm{PSF mag} \leq 19.0$), the TPR of level 2 selection is $0.948\pm{0.007}/0.995\pm{0.003}$ in $g$/$r$-band.


We also estimate the background rate after the level 2 selection using WF control data sample.
Among 33,339 alerts passing the level 1 selection, 11,622 alerts pass the level 2 selection. This corresponds to a background rate of $1.79\pm0.02$ alerts $\rm deg^{-2}$. 

\subsubsection{level 3 selection}
\label{sec:ana:EC2}
The level 3 selection is object-wise cut to remove remaining Galactic sources and AGNs. First, we basically follow the cut used in the ZTF filtering system: (1) if there is an object detected in Pan-STARRS1 (PS1) image within 1.5 arcsec, it should be a galaxy-like object based on “star-galaxy separation score" ({\tt sgscore}: \citealt{tachibana2018PASP..130l8001T}) ranging from 0 (galaxy-like) to 1 (star-like). While ZTF applies a criterion of ${\tt sgscore}<0.5$, we instead adopt ${\tt sgscore}<0.83$, following the FoM (Figure of Merit) threshold described in \cite{tachibana2018PASP..130l8001T}, in order to maintain a higher signal TPR. 
This threshold recover 99.5 \% of galaxy-like objects, which corresponds to a TPR of $0.9950 \pm{0.0004}$ in our filtering system. 

This relaxed cut leads to an increase in 
the number of background objects passing this criterion. 
Therefore, we add further criteria using cross match with catalog sources as follows: (2) No Gaia variable sources within 1.5 arcsec (\citealt{2023rimoldini, 2023eyer}, through Part 4 Variability catalog in \citealt{gaiavar2022yCat.1358....0G})\footnote{Gaia variable sources catalog is constructed by the data taken well before the detection of \triplet\ (from July 25, 2014, to May 28, 2017). Thus, they are unlikely to be the source that emit neutrino flare with the short time duration ($T_{\rm w}=30~{\rm days}$) considered in this study.}, (3) No Gaia sources with a significant ($> 3 \sigma$) proper motion within 1.5 arcsec (\citealt{gaiadr32023A&A...674A...1G}, through Part 1 Main source catalog in \citealt{gaiamain2022yCat.1355....0G}), and (4) No known AGNs within 1.5 arcsec \citep{2010veron, 2015flesch}. 

We estimate the fraction of our target lost by these cross matches. We use 9552 spectroscopically confirmed SNe from BTS. As a result, 12/14/26 SNe are removed by each criteria (Gaia variable/Gaia proper motion/AGN). From these, we estimate the TPR of each criterion as 
$0.9987\pm{0.0004}$, $0.9985\pm{0.0004}$, and $0.9973\pm{0.0005}$ for the criteria (2), (3), and (4) above, respectively.


We note this TPR is subject to uncertainties especially if the signal is a TDE. TDEs occur at the center of host galaxies while majority of SNe in the BTS samples do not occur at the center of the galaxies. Thus, our criteria above may filter out a fraction of TDEs that occur AGN-like host galaxy, which result in a reduction of TPR. We have checked that this effect is not significant by using the known TDE samples, but the number of TDE samples is still small (71 samples in BTS), preventing the estimate with a high statistics. Also, two out of three TDEs reported to be associated with the IceCube neutrino detections occurred in AGN hosts. Our criteria above may exclude such TDEs occuring in AGNs.

In our background analysis with WF control data sample, 397 objects pass the level 3 selection which corresponds to a background rate of $0.061\pm0.003$ objects $\rm deg^{-2}$.


\subsubsection{level 4 selection}
\label{sec:ana:EC3}

Finally, we apply the level 4 selection, object-wise light curve cut ($\Delta t_{\rm neg} \leq 50~\rm{days}$) to further remove remaining variable stars and AGNs. For this criterion, we use light curve of each object as additional information. 
For the light curve of each object obtained through the ZTF alert broker, ALeRCE API\footnote{https://alerce.readthedocs.io/en/latest/index.html},
we identify the significant ($> 5 \sigma$) negative detection.
Negative detections mean that the fluxes significantly decrease compared with the fluxes in the reference images.
We define $\Delta t_{\rm neg}$ as the time interval from the first to the last significant negative detection.
This criterion of $\Delta t_{\rm neg} \leq 50$ days effectively remove persistent objects such as variable stars and AGNs without removing SNe and TDEs. 

For this criterion, we estimate the TPR using historical light curve data of BTS spectroscopically confirmed SNe. We use 891 SNe located within $z\leq0.02$ from BTS samples with  well-sampled light curves. 
Before the TPR calculation, we remove 27 SNe with the reference images contaminated by SN fluxes. Among the remaining 864 SNe, 
only 2 SNe show the light curve with $\Delta t_{\rm neg} > 50$ days. 
Thus, we estimate the TPR of the level 4 selection as $0.9977\pm{0.0016}$. 

The number of objects passing the level 4 selection is 138 in the WF control data sample, which corresponds to a background rate of $0.021\pm0.002$ objects ${\rm deg^{-2}}$.

Finally, we compare the TPR of level 3 and 4 selections with the corresponding TPR of the ZTF standard filtering system. In the ZTF standard filtering system, a stringent criterion of {\tt sgscore} (${\tt sgscore}<0.5$ for PS1 objects within 1.5 arcsec) is imposed. To estimate the TPR of this criterion, we use transients reported in BTS between August 2019 and July 2020, which were selected using a looser condition of ${\tt sgscore}<0.76$ for PS1 objects within 2 arcsec \citep{perley2020ApJ...904...35P}. This estimate gives a TPR of 0.949. Note that this TPR should be regarded as an upper limit, since the true parent sample of transients would be larger.

On the other hand, in our filtering system, we impose a loose condition of {\tt sgscore} (${\tt sgscore}<0.83$), whose TPR is robustly estimated to be 0.995 by \citet{tachibana2018PASP..130l8001T}. An increase of background objects by this loose condition is further suppressed by additional criteria in level 3 and level 4 selections. The total TPR of our level 3 and 4 criteria is 0.987 ($=0.995\times0.9987\times0.9985\times0.9973\times0.9977$, see Table \ref{tab:tpr-nbg}). This demonstrates that our filtering system achieves a higher TPR than the ZTF standard filtering system.

\subsection{Multi-epoch analysis}
\label{sec:ana:multi}

The transient source search for \triplet~is facilitated by the ZTF observations at multiple epochs. When any single observation meets the detection-wise criteria, level 1 and 2 criteria are fulfilled. The incorporation of additional epochs into the counterpart search process enhances the resultant TPR, but it also raises the probability of contamination of background events (see also Appendix \ref{ap1} and Figure~\ref{fig:tpr-nbg}). It is imperative to judiciously optimize the number of epochs contributing to the search to ensure the efficacy of the methodology.

\begin{figure}[t!]
\epsscale{1.1}
\plotone{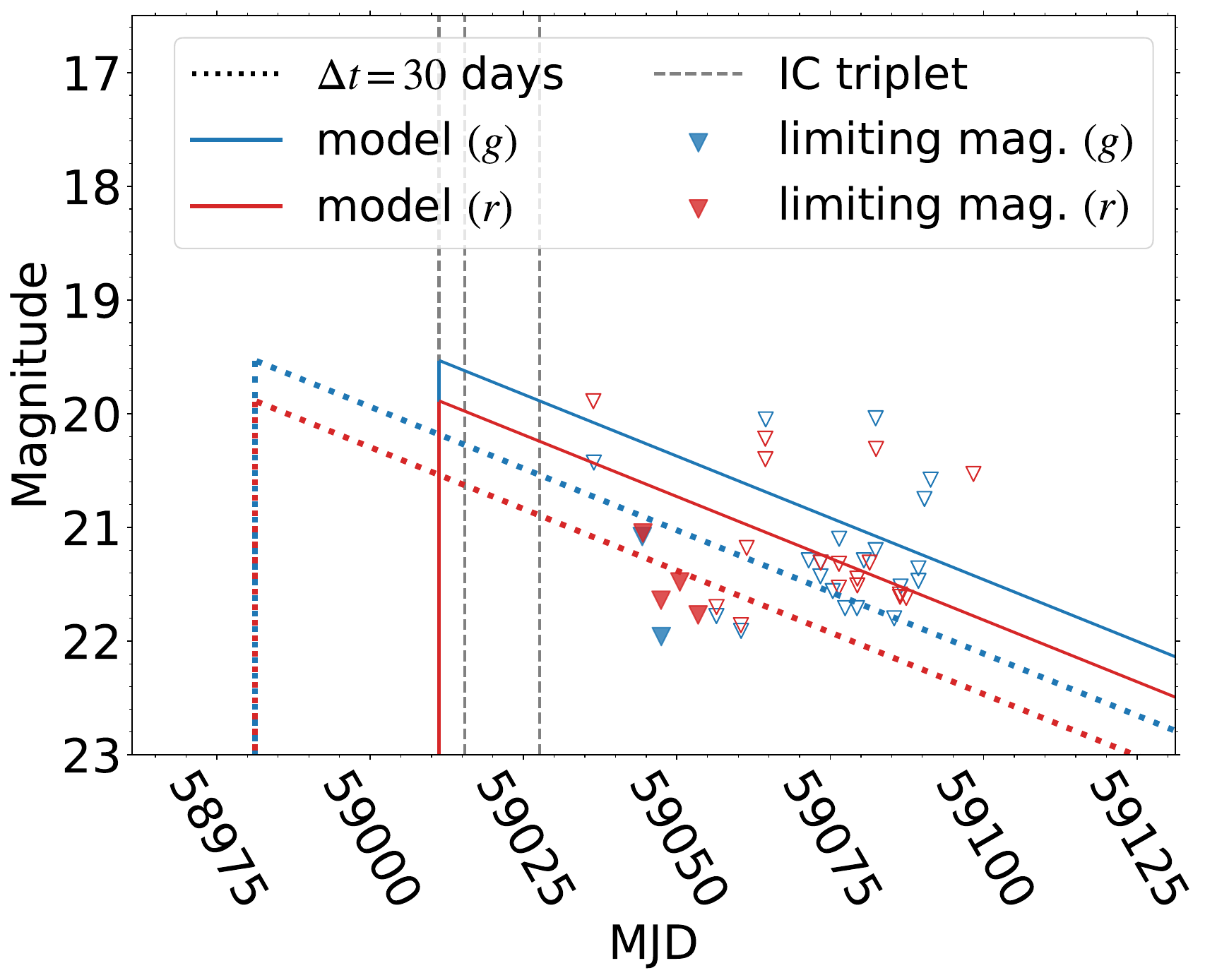}
\caption{Limiting magnitude of ZTF observations in each epoch at the IceCube localization area around the date of the neutrino detections (dashed lines). Filled points represent the epoch we used for multi-epoch analysis. Examples of our  light curve model are also shown with $L_{\rm pk}=10^{43}~{\rm erg~s^{-1}}$ and $t_{\rm decay} = 50~{\rm days}$ at redshift $z=0.05$. We assume black body SED with $T=20,000~\rm{K}$. Solid lines represent $\Delta t=0$ days and dotted lines represent $\Delta t = 30$ days. 
}
\label{fig:maglim}
\end{figure}

We trace the epoch dependence of TPR and the number of background events by applying the level 1 and level 2 selections to the ME1 control data sample, which contains 11 observing epochs. As we mentioned in Section \ref{sec:data:ztf}, we do not include the shallow observations on July 6, 2020, since the data on that day are not effective to detect dimmer objects and just increase the number of background objects. Each of the observation epochs has a different depth due to the night sky conditions as shown in Figure \ref{fig:maglim}, and we search for the best combination of epochs participating in the source search. The specific 6 epochs have been chosen to give an excellent TPR
($0.983$ in typical magnitude range, 18.5--19.0 mag) while suppressing the number of penetrating background 
events ($\lesssim 0.1~{\rm deg^2}$). 
We then apply the selections to the ME2 control data sample, which contains exactly the 6 selected epochs, but data from larger areas of the sky to gain statistical power.

The cumulative number of background objects is 67 objects, which corresponds to a background rate of $0.084\pm0.010~{\rm objects~deg^{-2}}$. For comparison, applying the ZTF standard filtering results in a background rate of $0.175\pm0.015~{\rm objects~deg^{-2}}$. The IceCube localization area ($3\sigma$) covered by the ZTF observation is 0.41 $\rm deg^2$ (Figure \ref{fig:localization}). Therefore, the background rate in the final analysis for the localization area of \triplet\ is $0.034\pm0.004$.

\begin{figure}[t!]
\epsscale{1.1}
\plotone{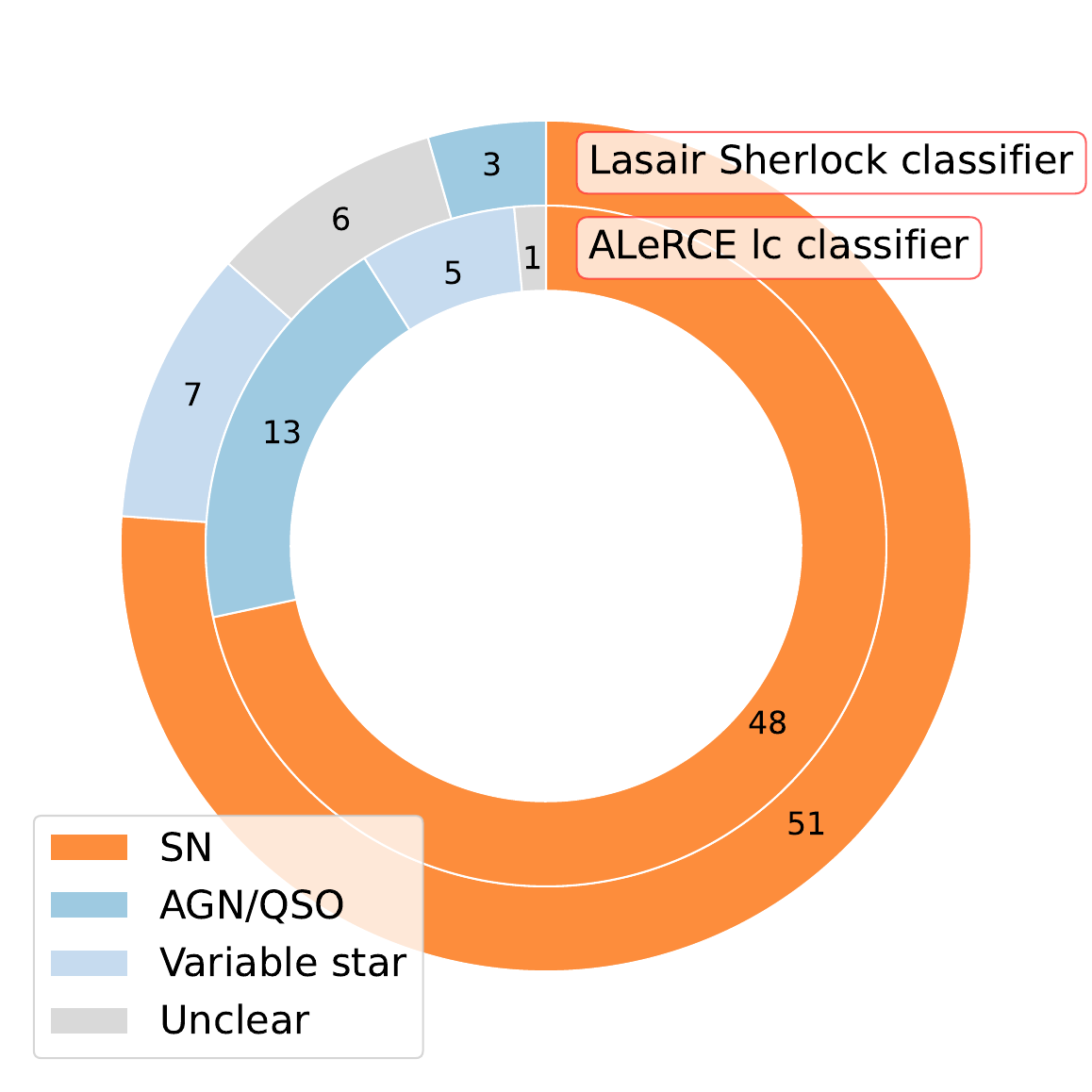}
\caption{Breakdown of objects identified in background anaysis with 6 epochs based on classifiers from ZTF alert broakers (ALeRCE light curve classifier and Lasair sherlock classifier).}
\label{fig:breakdown}
\end{figure}

We examine the nature of the background objects remaining in the final sample after applying all the cuts to the ME2 control data. They are dominated by true astrophysical sources.
Figure \ref{fig:breakdown} shows the breakdown of classification from ZTF brokers, ALeRCE light curve classifier~\citep{forster2021AJ....161..242F} and Lasair Sherlock classifier~\citep{2019smith}. Typically 70\% of background objects are SN-like objects including 13\% of spectroscopically confirmed SNe. Thus, a majority (70 \%) of the background objects are true transients, which cannot be further reduced. Note that the number of SNe is broadly consistent with the expectation from the typical SN rate and ZTF sensitivity. 


The dominance of SNe as background objects means that the background rate may vary depending on the direction, reflecting the variation in the number of SNe due to the inhomogeneous distribution of galaxies at low redshift.
This effect may be estimated by integrating the luminosity of the galaxies toward each direction.
The SN rate can be approximated by the total star formation rate, and the star formation rate is roughly proportional to the $B$-band luminosity 
\citep{1998kennicuttARA&A..36..189K}. 
By using the GLADE+ catalog~\citep{2022gladeMNRAS.514.1403D}, we estimate the integrated $B$-band luminosity ($L_{\rm B}$) of galaxies per unit area both for the region used for background analysis and the localization area of \triplet. 
We find that the $L_{\rm B}$ in the localization area is 1.26 times higher than the median of the background region.
The standard deviation of $L_{\rm B}$ in the background regions is 0.089 dex.
For a conservative estimation of the background, we define a correction factor of 1.55, which is the ratio between $L_{\rm B}$ in the localization area and $L_{\rm B}$ at $1\sigma$ below the median value of the background regions.
We scale the background rate by this factor, resulting in the final background rate of $0.053\pm0.006$, which is used in the signal model test \footnote{Most of background SNe are located at $z=0.05-0.1$ while GLADE+ catalog is highly incomplete at such a redshift range. 
Thus, the derived variation of $L_{\rm B}$ in the background region is overestimated due to a high fraction of nearby galaxies. Thus, the correction factor by using $L_{\rm B}$ at 1 $\sigma$ below the median value gives a very conservative estimate of the background rate.
}.

\subsection{Signal model tests}
\label{sec:ana:test}

We discuss our statistical analysis method to test our signal model. Here, we assume that the triplet is of astrophysical origin and set the signalness to unity. We look for transient objects in the neutrino localization area, and the observable is the number of the transients that pass through our event selections,  $n_T$ (see Section \ref{sec:ana:selection}). Our signal hypothesis is that the neutrino-emitting object has a set of $(L_{\rm pk},~t_{\rm decay},~\Delta t)$.  

We estimate the probability that ZTF fails to detect the object at a distance $z$ at epochs $t_i$ as 
\begin{equation}
p_{\rm fail}(z)=\prod_{i=1}^6(1-{\rm TPR_{\rm 1}({\it t_i,z})TPR_{2}}(t_i,z)),
\end{equation}
where ${\rm TPR}_{n}$ denotes the TPR of $n$th selection process.
The sensitivity of the ZTF data depends on the epoch $t_i$.
Thus, when the flux by our signal model is lower than the sensitivity limit, 
 we set ${\rm TPR_{1}} (t_i, z)=0$
 \footnote{Since we choose the 6 epochs that have deep enough limiting magnitudes, this treatment does not affect our resulting constraint maps.}. In addition, we count the number of epochs that satisfy the condition that the signal flux is larger than the sky noise level, and set $p_{\rm fail}=1$ if the number count is less than or equal to 1. 
Since $\rm TPR_1$ depends on the optical flux at the epoch $t_i$, $p_{\rm fail}$ depends on the source parameters, i.e., a higher $L_{\rm pk}$, a longer $t_{\rm delay}$, and a shorter $\Delta t$ lead to a lower $p_{\rm fail}$.
Taking into accout the source redshift distribution, we estimate the probability that ZTF detects the object to be
\begin{equation}
n_{\rm sig} = {\rm TPR}_{\rm obj}f_{\rm cover}\int dz(1-p_{\rm fail}(z)) P_{\rm dist}(z) ,
\end{equation}
where $f_{\rm cover}$ is the coverage of the neutrino error region and ${\rm TPR_{obj} = TPR_{3}TPR_{4}}$ is the TPR of object-wise selections (levels 3 and 4). $P_{\rm dist}(z)$ is the probability distribution functions of the sources of triplet events introduced in Section \ref{sec:ana:model}.
We should note that $0<n_{\rm sig}<1$ is always satisfied. Given the background event, $\mu_B$, which should follow the Poisson distribution,
we define likelihood as 
\begin{eqnarray}
    \mathcal L =\left\{ 
    \begin{array}{ll}
    e^{-\mu_B}(1-n_{\rm sig})~~&~~(n_T=0)\\
    1 - e^{-\mu_B}(1-n_{\rm sig})~~ &~~(n_T\ge1)
    \end{array}
    \right.,
\end{eqnarray}
where $n_T$ is the number of transients found in the localization area.
Our signal hypothesis is compared with an alternative hypothesis in which the neutrino source has a different set of parameters. The test statistic is given as 
\begin{equation}
\Lambda=\ln \frac{\mathcal L(L_{\rm pk},t_{\rm decay}~\Delta t)}{\mathcal L(\hat L_{\rm pk},~\hat t_{\rm decay}~,\hat{\Delta t})},
\end{equation}
where $(\hat L_{\rm pk},~\hat t_{\rm decay}~,\hat{\Delta t})$ are chosen so that $\mathcal L$ is maximized. 

We perform mock observations for various sets of $(L_{\rm pk},~t_{\rm decay},~\Delta t)$ to obtain the distribution of $\Lambda$. We perform $10^4$ mock observations for a given parameter set and obtain 68\% and 90\% confidence area according to Feldman \& Cousins method \citep{feldman1998PhRvD..57.3873F}.

\section{Results} \label{sec:result}

\subsection{Unblinding Data in the Localization Area}

We apply our selection criteria to the data in the localization area from July 14 to 23, 2020. As a result, no candidates in the localization area pass the level 2 selection. Therefore, we place the constraints on the parameters of transients based on non-detection ($n_T=0$).

\subsection{Parameter Constraints}

\begin{figure*}[t!]
\epsscale{1.0}
\begin{minipage}[l]{0.48\hsize}
 \begin{center}
\includegraphics[width=\linewidth]{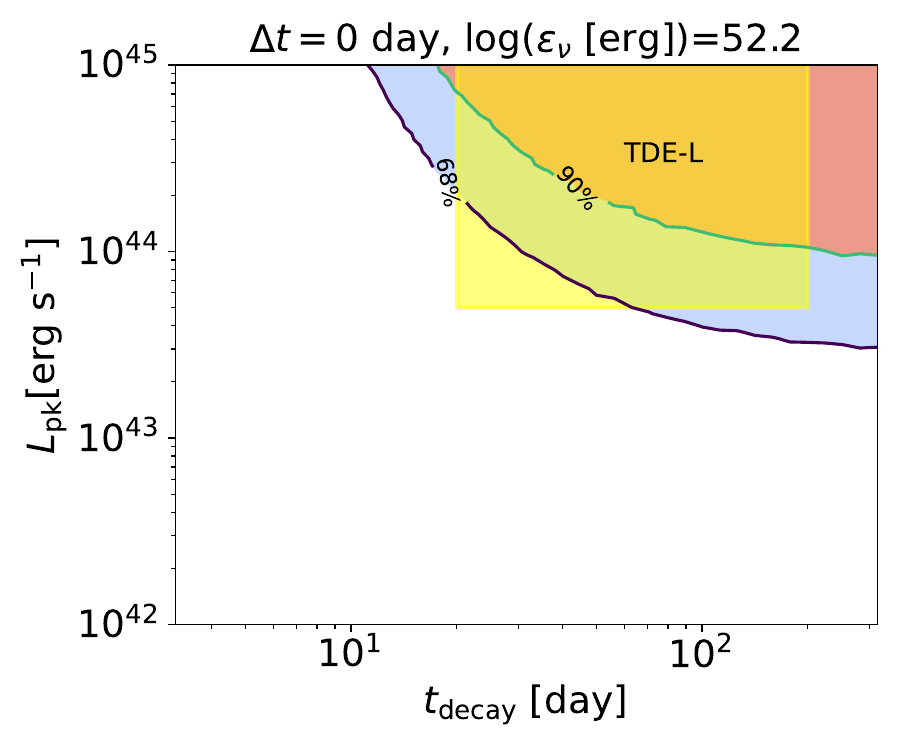}
\includegraphics[width=\linewidth]{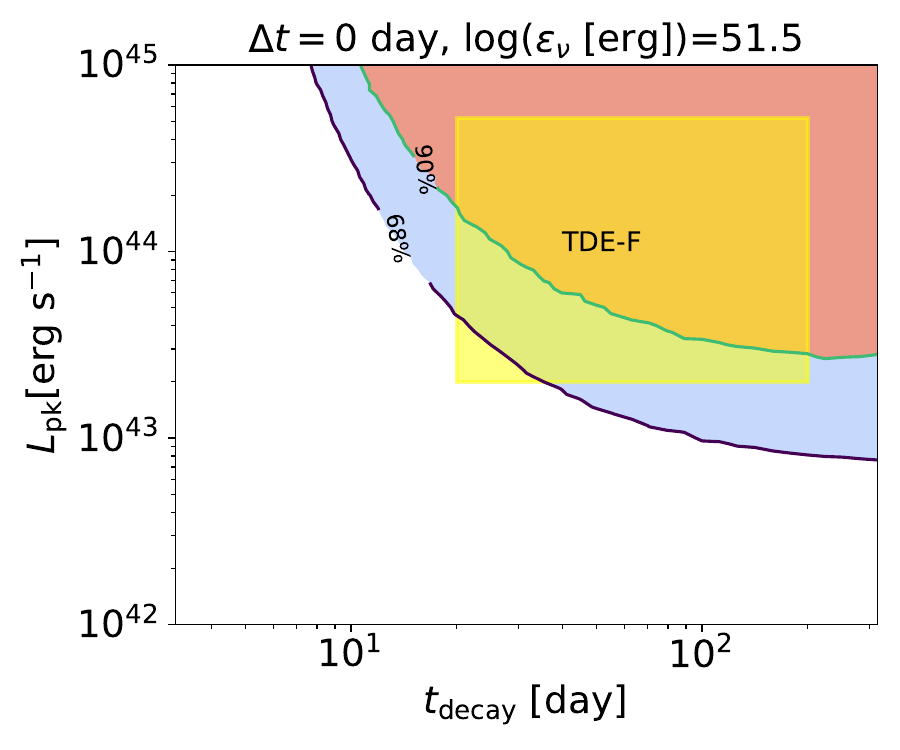}
\includegraphics[width=\linewidth]{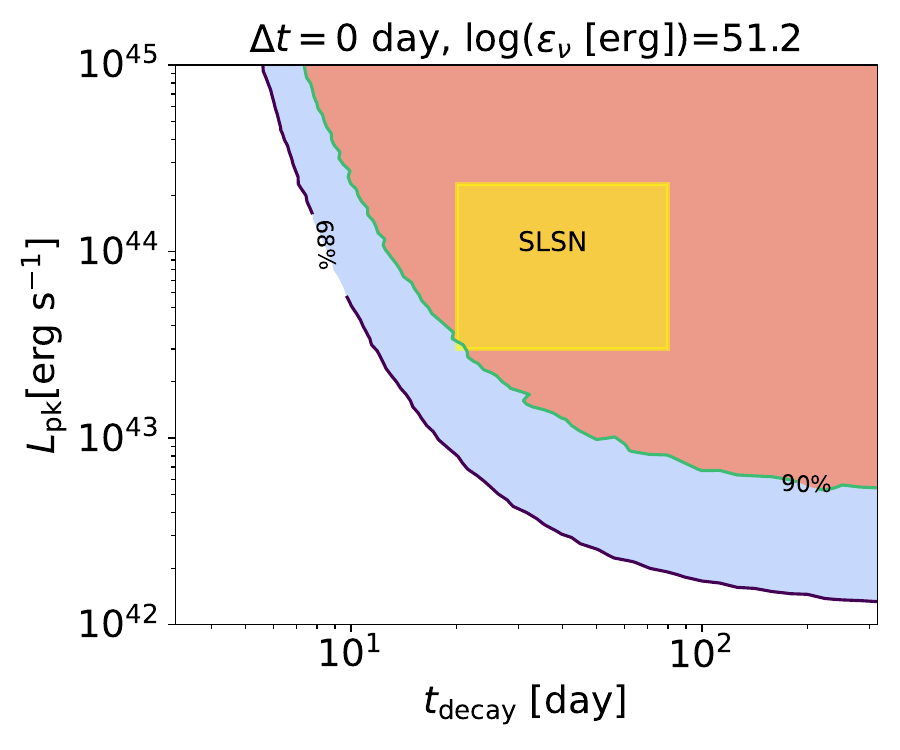}
\end{center}
\end{minipage}
\begin{minipage}[l]{0.48\hsize}
 \begin{center}
\includegraphics[width=\linewidth]{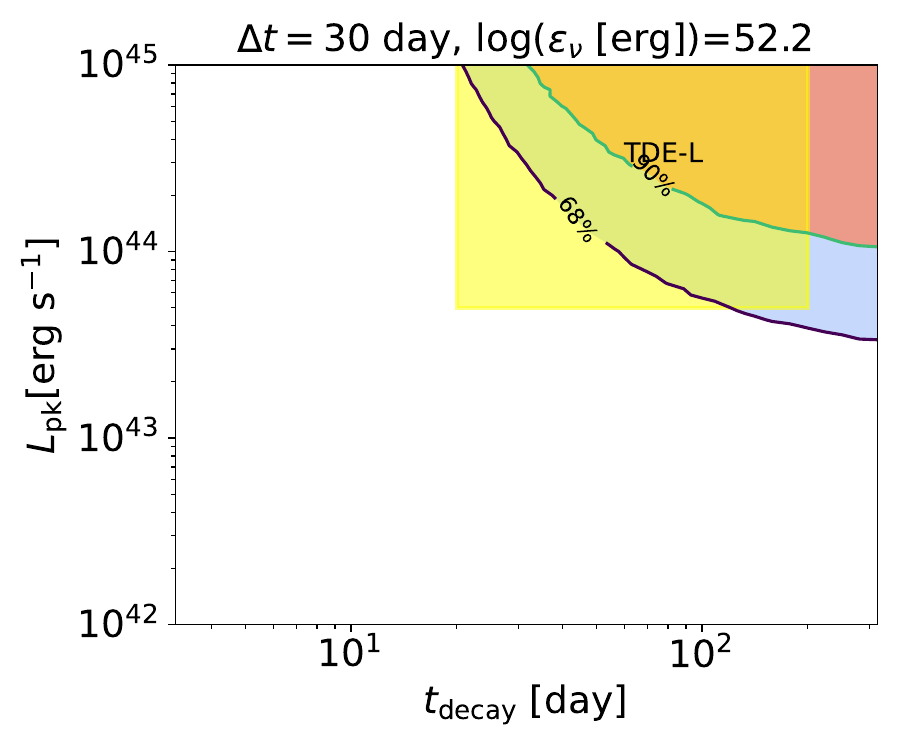}
\includegraphics[width=\linewidth]{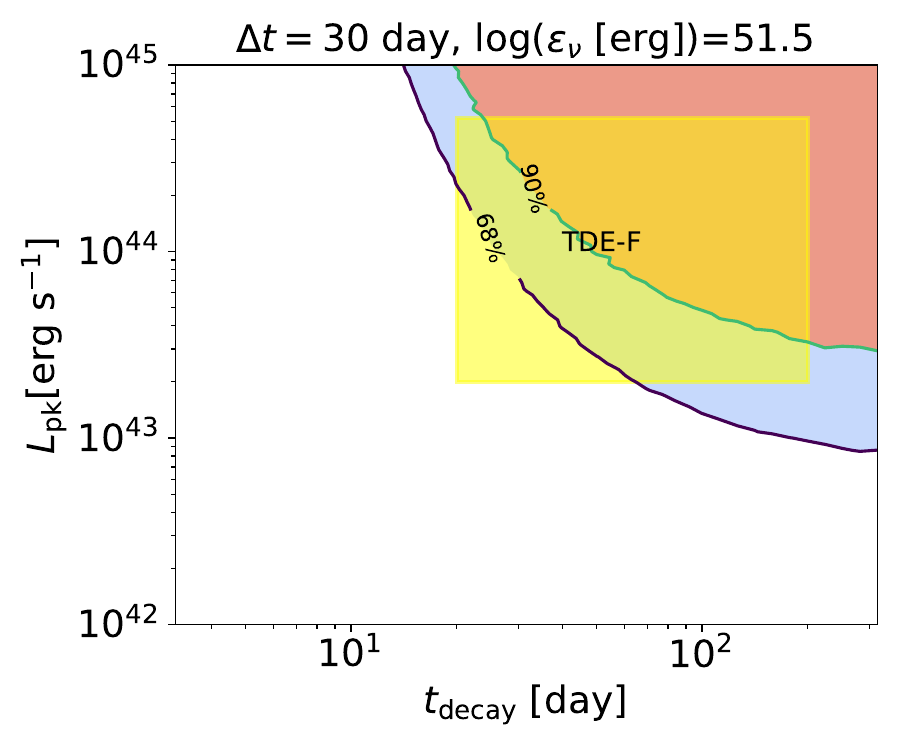}
\includegraphics[width=\linewidth]{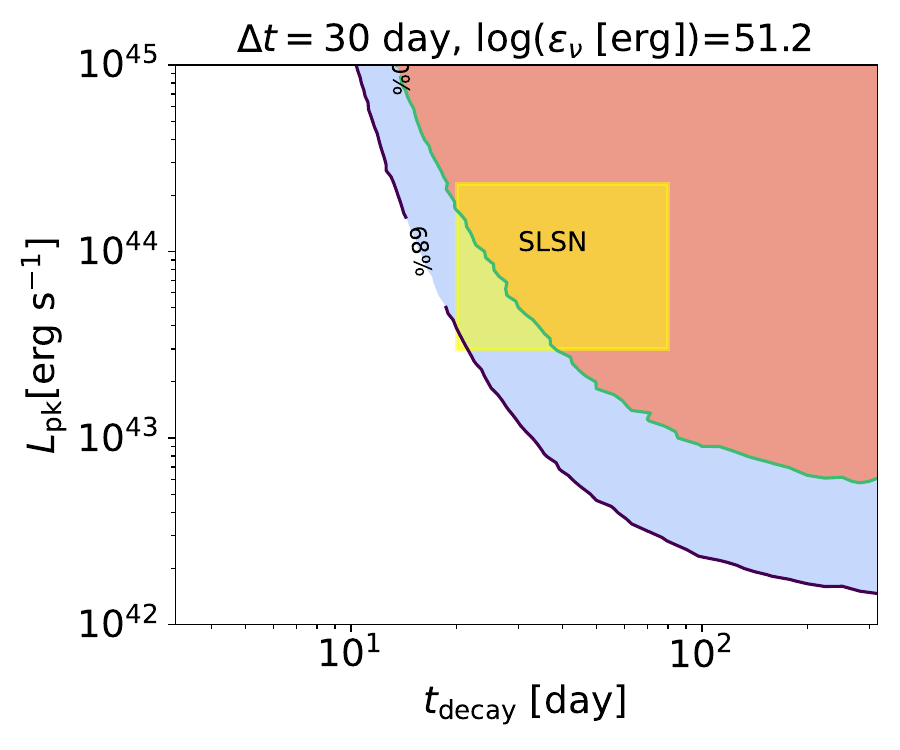}
\end{center}
\end{minipage}
\caption{Constraints in $t_{\rm decay}-L_{\rm pk}$ plane for TDE-L (top), TDE-F (middle), and SLSN (bottom) with $\Delta t=t_{\nu, 1}-t_{\rm pk} = 0$ (left) and 30 days (right). Orange and cyan regions are constrained by 90\% and 68\% confidence level. Yellow regions are typical parameter spaces for TDEs \citep{yao2023ApJ...955L...6Y} and SLSNe \citep{moriya2024arXiv240712302M}. } 
\label{fig:contour}
\end{figure*}

We calculate the test statistic with $n_T=0$ and put constraints on $L_{\rm pk}$ and $t_{\rm decay}$ for two cases with $\Delta t=0$ and 30 days.
Figure \ref{fig:contour} shows the region where the parameters are disfavored with significance levels of 68 \% and 90 \%. 
In general, as expected, a parameter space with a high $L_{\rm pk}$ and a long $t_{\rm decay}$ is more strongly constrained as such objects are more easily detectable by ZTF. 

For a lower value of $\varepsilon_\nu$, it is expected that the neutrino source is located at a nearby distance (Figure \ref{fig:zpdf}), resulting in a better constraint. 
This is the reason why a wider parameter space is constrained for the TDE-F and SLSN cases as compared with TDE-L case.
As for the timing of the peak, a smaller $\Delta t$ leads to brighter optical signals at the time of ZTF observations.
Thus, the constrained parameter space is wider for the case with $\Delta t = 0$ compared with the case with $\Delta t = 30$ days.

For the cases with $\Delta t=0$ (left panels in Figure \ref{fig:contour}), the constrained regions cover almost all ranges of observed TDEs and SLSNe. 
For $\Delta t=30$ days, almost all the observed SLSNe parameters are constrained, but faint and short-duration TDEs are not constrained for TDE-L. 
Nevertheless, TDE-L case is as meaningful as TDE-F case: TDE-L assumes a lower $R_0$, which implies that relatively luminous TDEs contribute to the cosmic neutrino background. Thus, we expect more luminous TDEs for the case with TDE-L.

\section{Discussion} \label{sec:discussion}


\subsection{Implications for TDE models}

TDEs have been discussed as high-energy neutrino sources since the discovery of the jetted TDEs in 2011 \citep{wang2011PhRvD..84h1301W,lunardini2017PhRvD..95l3001L,dai2017MNRAS.469.1354D}. However, it turned out that they are too rare to be the dominant source of cosmic neutrino background observed on Earth; Non-detection of IceCube lower-energy neutrinos constrain the rare population of neutrino sources with $R_0\lesssim10^{-7}\rm~Mpc^{-3}~yr^{-1}$ \citep{icecube2023ApJ...951...45A}, which strongly disfavors the jetted TDEs ($R_0\lesssim 3\times10^{-10}\rm~Mpc^{-3}~yr^{-1}$; \citealt{brown2015MNRAS.452.4297B}).

Our analysis disfavors long-duration transients as neutrino sources, especially for $t_{\rm decay}\gtrsim 50$ days. However, this does not mean that we could disfavor TDEs as the dominant source of the cosmic neutrino background.
All the candidate TDE-neutrino associations reported so far are non-jetted TDEs and have long time lags between the optical peak and neutrino detection, typically $\gtrsim$ 100 days \citep{stein2021,reusch2022,2023jiang,vanvelzen2024MNRAS.529.2559V,2024yuan}. 
This feature could be explained by some of current scenarios, such as collisionless corona formation \citep{murase2020ApJ...902..108M}, wind-cloud interactions \citep{wu2022MNRAS.514.4406W}, choked delayed jets \citep{zheng2023ApJ...954...17Z,mukhopadhyay2024MNRAS.534.1528M}, and IR dust echos \citep{winter2023}. 
If such a delayed neutrino emission is typical, a TDE is unlikely to be the origin of triplet events with a 30-day time window, as the expected duration of the neutrino emission is much longer than the time window of the multiplet alert. In this case, we cannot constrain the TDE-neutrino paradigm by this analysis. 


Nevertheless, TDEs may produce triplet signal within 30 days if the optical and neutrino luminosity of TDE evolve rapidly (under the relation of $L_\nu \propto L_{\rm opt}$). Such TDEs are observed by recent optical transient surveys \citep{yao2023ApJ...955L...6Y, Hammerstein2023ApJ...942....9H}, although association with a neutrino signal is not reported.
These may be realized when black-hole mass is low based on magnetized accretion disk scenario \citep{hayasaki2019ApJ...886..114H,murase2020ApJ...902..108M} or if wind dissipation occurs in a smaller radii based on outflow scenarios \citep{zhang2017PhRvD..96f3007Z,murase2020ApJ...902..108M}. 
It is challenging to identify rapidly evolving TDEs, especially for $t_{\rm decay}\lesssim20$ days, for the case of \triplet\ because there is a time lag between the triplet event and the available ZTF data. 
For future multiplet events, quick follow-up observations within a few days are desirable to identify relatively fast evolving transients of $t_{\rm decay} \sim 10-30$ days. 

To constrain the TDE-neutrino paradigm, quick UV follow-up observations are also essential. Non-negligible fraction of non-jetted TDEs detected in optical bands have high temperature of $T> 2\times10^4$ K \citep{yao2023ApJ...955L...6Y}. The peak frequency of the photon spectrum for such high-temperature transients exceeds the optical band, leading to a lower detectability in $g$- and $r$-bands. UV observations, which will be available with Rubin/LSST \citep{ivezic2019}, ULTRASAT \citep{Rhoads2024SPIE13093E..35R}, and PETREL \citep{yatsu2024SPIE13093E..0FY} near future, will help improve the detectability of the higher temperature transients.

\subsection{Implications for SLSN models}

Although SLSNe have not been discussed as high-energy neutrino sources frequently, they are good candidates of high-energy neutrino sources (See \citealt{pitik2022ApJ...929..163P} for possible association between IC-200530 \& AT2019fdr as a SLSN. However, AT2019fdr is also considered to be a TDE; cf.~\citealt{reusch2022}.). The power source of SLSNe are still uncertain \citep{moriya2024arXiv240712302M}, and three models are actively discussed: radioactive decay of a large amount of $^{56}$Ni, CSM interaction, and energy injection by a central remnant. Although high-energy neutrinos are unlikely to be produced for the radioactive-decay scenario, the other two power-sources have been actively discussed as high-energy neutrino sources in the context of core-collapse SNe \citep{2011murase,zirakashvili2016APh....78...28Z,petropoulou2017MNRAS.470.1881P,fang2020ApJ...904....4F,kimura2024arXiv240918935K} and neutron-star mergers \citep{fang2017ApJ...849..153F,muchopadhyay2024arXiv240704767M}. Both of these scenarios predict that the time window for neutrino emission is about $10-30$ days, which is a good match with the multiplet alert. 

The main differences of neutrino signals in these two models are their expected neutrino spectra: CSM interaction SNe produce neutrinos via $pp$ channel with shock-accelerated CRs, which leads to the canonical spectra of $dN/dE_\nu \sim E_\nu^{-2}$ or softer \citep[e.g.][]{kimura2024arXiv240918935K}. On the other hand, magnetar engines produce neutrinos via the $p\gamma$ channel with CRs accelerated at polar-cap or reconnection, which make the neutrino spectra much harder than the canonical one, typically $dN/dE_\nu \sim E_\nu^{-1}$ with the spectral peak at EeV energies \citep[e.g.][]{muchopadhyay2024arXiv240704767M}. Considering these features, central-engine SLSNe are likely to produce multiplet signals with PeV-EeV energies. Thus, the current multiplet alert with TeV-PeV energies should be most likely associated with the interaction-powered SLSNe. 

Our null result disfavors SLSNe as the dominant source of cosmic neutrino background. This is consistent with the theoretical expectation based on an energetics argument shown below. CSM interaction features are observed not only SLSNe but also SNe IIn that are roughly 100 times more frequent but 10 times less luminous. This implies that SN IIn have 10 times higher energy budget than SLSNe, and thus, the cosmic neutrino background should be dominated by SN IIn as long as both SN IIn and SLSNe share the same energy release mechanism through the CSM interaction (cf. \citealt{icecubeiin2023}). Although current neutrino detectors are not sensitive enough to detect multiplet signals from SN IIn, multiplet alerts with future detectors would be able to constrain this type of transients as cosmic neutrino sources.

\subsection{Prospects}

Compared to singlet alerts, multiplet alerts have a much better ability to identify transient neutrino sources. Neutrino sources of singlet alerts are located in a cosmological distance (see Figure \ref{fig:zpdf}), which can be detectable only with instruments with deep and wide imaging capabilities, such as Blanco/DECam~\citep{flaugher2015}, Subaru/HSC \citep{miyazaki2018}, and Rubin/LSST \citep{ivezic2019}. 
\cite{2019morgan} discusses future prospects of detecting core-collapse SNe as neutrino sources using singlet alerts with DECam. They concluded that $\sim2\sigma$ detection requires $\sim60$ follow-up observations, even if the follow-up observations are performed to neutrino alerts with signalness equal to unity (signalness is the probability that the neutrino event is astrophysical origin). In addition, singlet alerts in general could originate from steady sources that do not show strong variability, such as Seyfert galaxies and cosmic-ray reservoirs. If this is the case, we cannot identify neutrino sources by follow-up observations. 

In contrast, multiplet alerts enable us to focus on nearby transients as neutrino sources. Since we have a time window of $T_\mathrm{w} = $ 30~days, it is highly unlikely to be produced by weakly variable steady sources discussed above. Also, multiplet sources need to be located within $z\lesssim0.3$ (see Figure \ref{fig:zpdf}), allowing us to detect it with transient surveys with smaller aperture telescopes, such as ATLAS \citep{tonry2018}, ZTF \citep{bellm2019PASP..131a8002B}, and PS1 \citep{chambers2016}.
As demonstrated in the previous section, just one follow-up observation will provide a good constraint, or $\sim1.5-2\sigma$ detection if an optical transient of interest is discovered. To obtain a better constraint or detect rapidly evolving transients, quick optical follow-up observations to multiplet alerts are essential. Although the signalness of multiplet alerts could be typically lower than high-quality singlet alerts, neutrino source searches using smaller telescopes with multiplet alerts are highly beneficial.




\section{Summary} \label{sec:summary}


We conduct archival search for the optical counterpart of the IceCube triplet event \triplet\ detected in 2020. By using ZTF data, we develop our alert filtering system and validate the performance in a blind analysis strategy.
Our filtering system achieves a signal $\rm TPR= 0.983$ for transients in typical magnitude range for the ZTF (18.5--19.0 mag) and background rate $0.084~\rm{objects~deg^{-2}}$ with 6 epochs observations. Applying this filtering system to the actual data in the localization area, we find no transient candidates within the localization area from the data in 6 epochs (July 14 to 23, 2020), about one month after the detections of the IceCube triplet event.

Assuming that the IceCube triplet event originates from an astrophysical source, we calculate the test statistics to constrain the parameters of optical light curve $L_{\rm pk}$ and $t_{\rm decay}$ using a simple signal model that assumes optical signals from TDEs and SLSNe. 
For the cases of $\Delta t=0$ (no time lag between neutrino detections and optical peak), the constrained regions cover almost entire range of the observed TDEs and SLSNe. 
Although we should be cautious as the signalness of the multiplet event could be typically lower, our study demonstrates that follow-up of just one triplet event can give strong constraints on the parameters of transients.

In this study, we could not constrain transients with short duration $t_{\rm decay} \leq30~\rm{days}$ 
due to the lack of optical data just after the detections of the triplet. 
For future follow-up observations of IceCube multiplet events, quick follow-up observations are desirable to testify transients with a fast evolving light curve. 

\begin{acknowledgments}


This work is supported by the Grant-in-Aid for Scientific research from JSPS for Transformative Research Areas (A) (grant No. 23H04891, 23H04892, 23H04894, 23H04899). S.T. acknowledges support from Graduate Program on Physics for the Universe (GP-PU) at Tohoku University.

The presented study is based on observations obtained with the Samuel Oschin Telescope 48-inch and the 60-inch Telescope at the Palomar Observatory as part of the Zwicky Transient Facility project. The ZTF is supported by the National Science Foundation under Grants No. AST-1440341 and AST-2034437 and a collaboration including current partners Caltech, IPAC, the Oskar Klein Center at Stockholm University, the University of Maryland, University of California, Berkeley, the University of Wisconsin at Milwaukee, University of Warwick, Ruhr University, Cornell University, Northwestern University and Drexel University. Operations are conducted by COO, IPAC, and UW.

This work has made use of data from the European Space Agency (ESA) mission
{\it Gaia} (\url{https://www.cosmos.esa.int/gaia}), processed by the {\it Gaia}
Data Processing and Analysis Consortium (DPAC,
\url{https://www.cosmos.esa.int/web/gaia/dpac/consortium}). Funding for the DPAC
has been provided by national institutions, in particular the institutions
participating in the {\it Gaia} Multilateral Agreement.

\end{acknowledgments}





\appendix

\section{Signalness of neutrino multiplets}
\label{ap-signalness}

The signalness of neutrino multiplet is defined as a fraction of multiplets originating from astrophysical transient sources to the total number of multiplets with identical test statistics, $\Lambda$ (see \cite{2025shimizu} for the derivation of $\Lambda$). 
The distribution of $\Lambda$ from astrophysical transients, $dN^\mathrm{astro}/d\Lambda$, is computed assuming an emission energy of neutrinos per a source, $\varepsilon_\nu $ and an event rate of the transients $R_0$ as:\vspace{0mm}
\begin{align}
\frac{dN^\mathrm{astro}}{d\Lambda} = T_\mathrm{obs}R_0 \int dz\, \psi(z) \,\rho(\Lambda;\mu(\varepsilon_\nu ,z)) \frac{dV}{dz} ,
\end{align}
where $T_\mathrm{obs}$ is a total livetime of a dataset, ${dV}$ is a volume element, $z$ is a redshift of the source, and $\psi(z)$ is an evolution of sources with redshift which is compatible with the star formation rate~\cite{2025shimizu}. A factor of $\rho(\Lambda;\mu(\varepsilon_\nu ))$ represents a normalized distribution of $\Lambda$ ({\it i.e.,} a probability density function) when the averaged number of neutrino detection in the multiplet time window (30~days) is $\mu$, and is evaluated by the simulation of injections of signal neutrino events. The value of $\mu$ is a function of $\varepsilon_\nu$ and $z$, 
and computed based on the neutrino flux from a source at $z$ and an effective detection area of the IceCube. The distribution of $\Lambda$ from backgrounds (atmospheric neutrinos and unrelated astrophysical neutrinos) in the livetime of $T_\mathrm{obs}$, $dN^\mathrm{bg}/d\Lambda$, is evaluated by a data-driven background control sample. The signalness at $\Lambda$ is thus given by:
\begin{align}
\mathcal{S}(\Lambda)=\frac{\frac{dN^\mathrm{astro}}{d\Lambda}}{\frac{dN^\mathrm{astro}}{d\Lambda}+\frac{dN^\mathrm{bg}}{d\Lambda}}.
\end{align}
From Figure~4 (bottom panel) of \citealt{2025shimizu}, the largest test statistic 
value of triplet, which corresponds to \triplet, indicates $\mathcal{S}(\Lambda)\sim 60\%$.

\section{Background analysis}
\label{ap1}

\begin{figure}[t!]
\epsscale{1.1}
\plotone{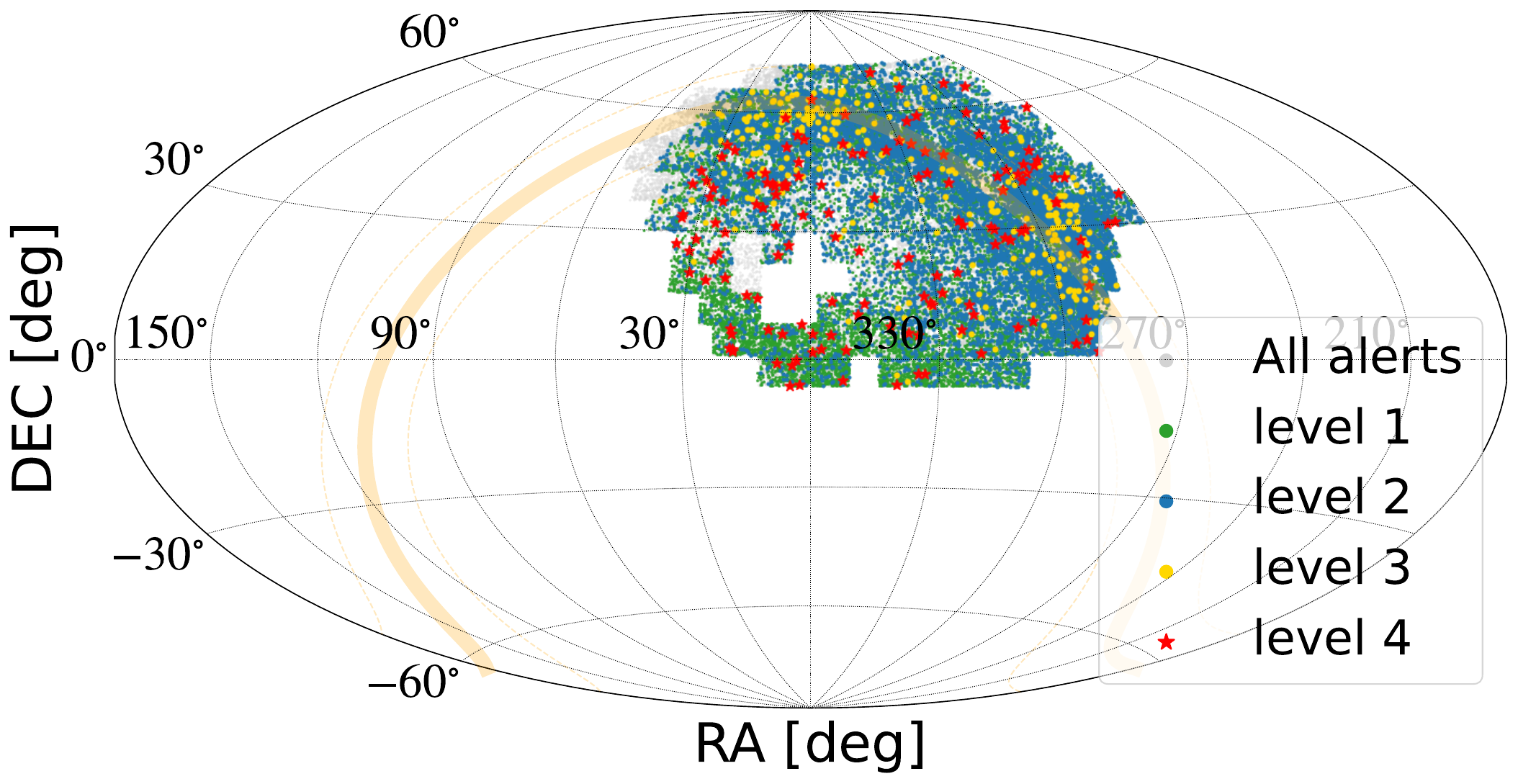}
\caption{Distribution of the alerts passing each criterion in WF control sample. Orange line represents the Galactic plane with orange dashed lines for galactic latitude $b=\pm10~\rm{deg}$.}
\label{fig:alerts}
\end{figure}

For the background modeling (Section~\ref{sec:ana:bg}), we use control data samples from three fields: the Wide Field (WF) sample, the Multiple Epoch 1 (ME1) sample, and the Multiple Epoch 2 (ME2) sample.
The WF sample consists of all alerts in the $g$-band on July 30, 2020, excluding the region around the direction of the IceCube triplet event \triplet.
Figure~\ref{fig:alerts} shows the sky distribution of the alerts of the WF sample passing each selection level. 

\begin{figure}[t!]
\epsscale{1.1}
\plotone{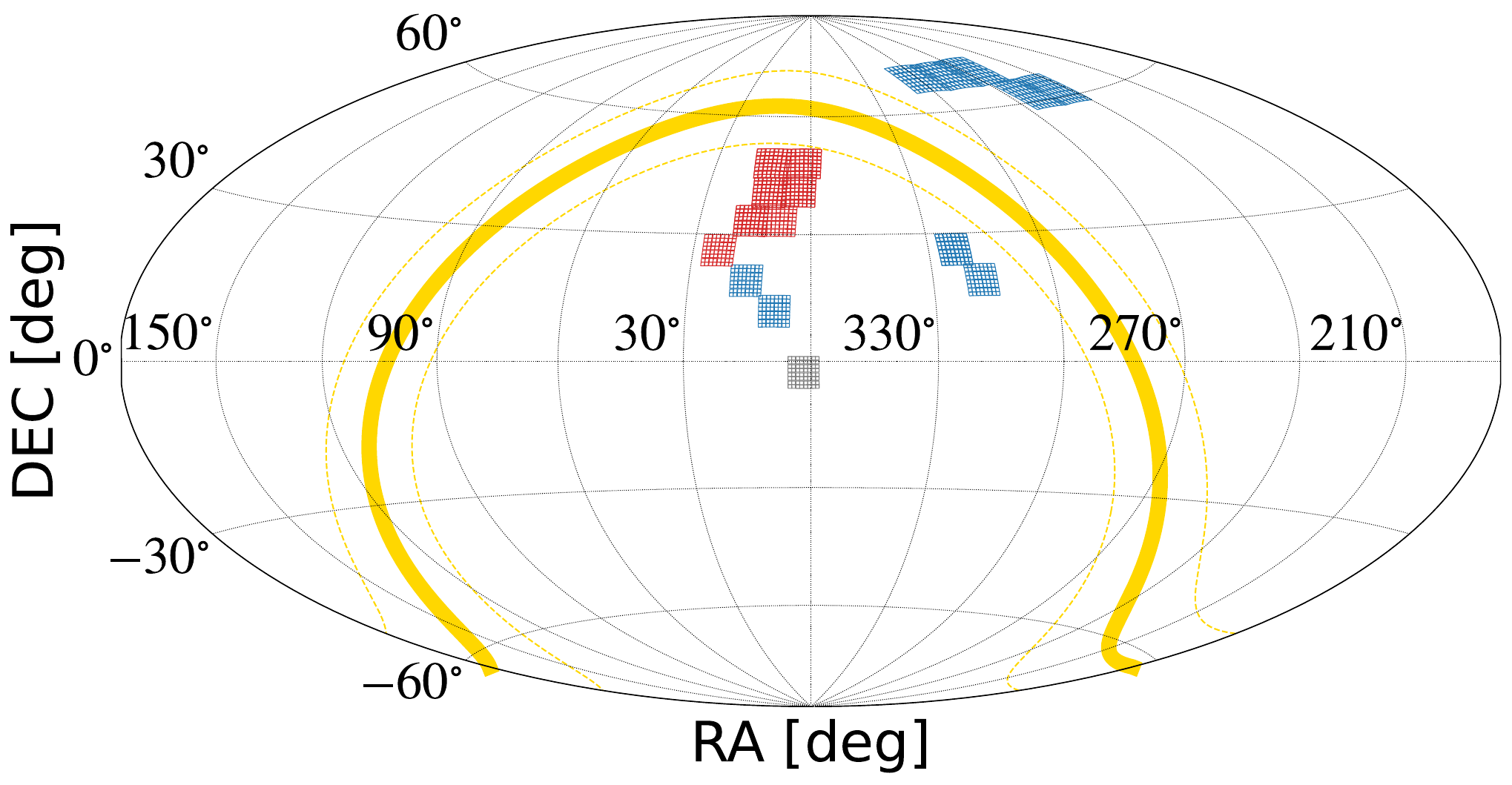}
\caption{The sky region used in our multi-epoch analysis. Red regions represent the area repeatedly observed from July 14 to 31, 2020. For 6 epochs analysis, we add blue regions which are repeatedly observed from July 14 to July 23. The gray region represents the field which includes IceCube localization area of the triplet \triplet.}
\label{fig:overlap}
\end{figure}

\begin{figure}[t!]
\epsscale{1.1}
\plotone{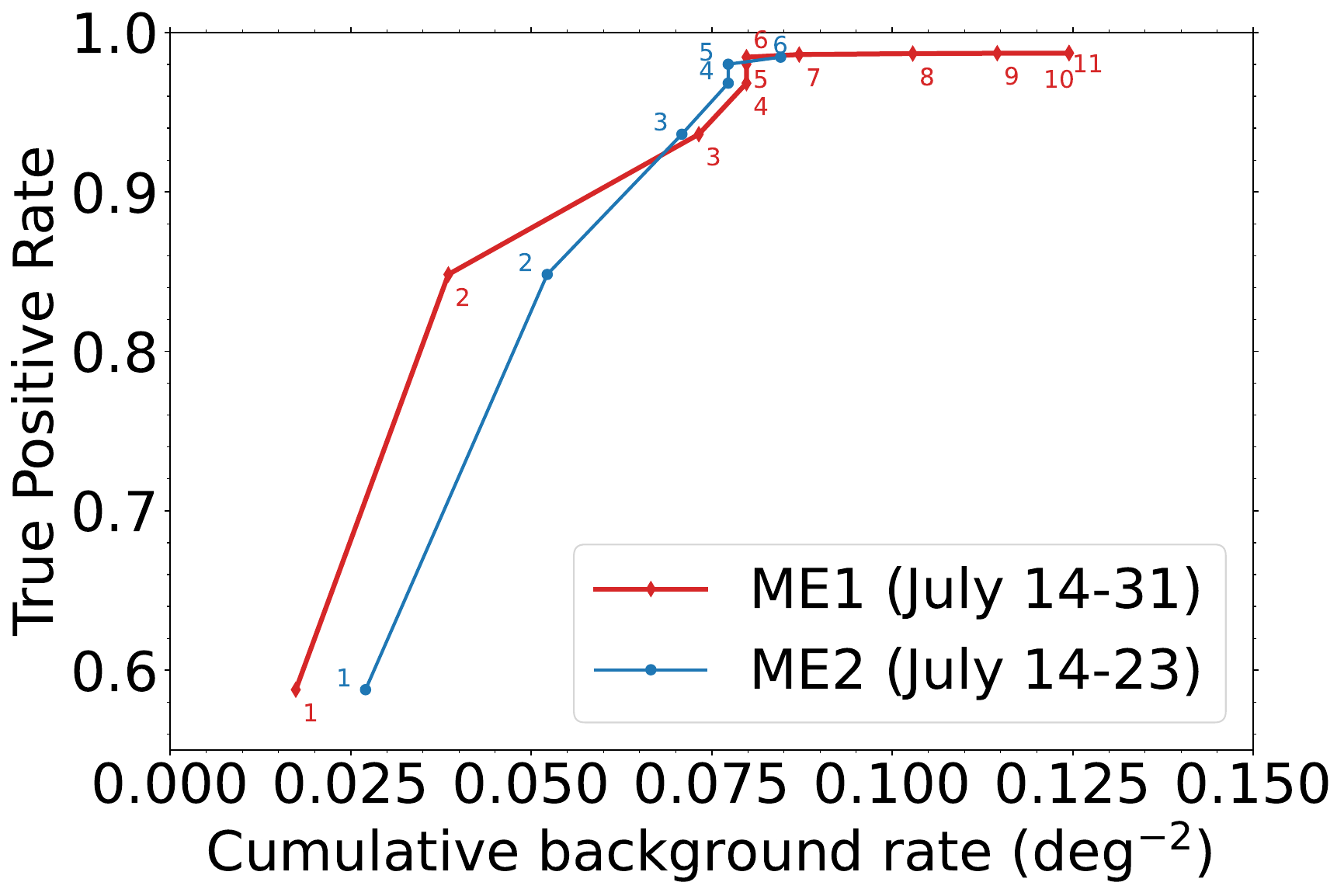}
\caption{Relations between the number of background objects and the TPR along with repeating observations. Red line represents measurement using the ME1 sample (July 14 to July 31) and blue line represents the measurement using the ME2 sample (July 14 to July 23) respectively. We assume that the objects are in the magnitude range of 18.5--19.0 mag to calculate the TPR.}
\label{fig:tpr-nbg}
\end{figure}

As described in Section~\ref{sec:ana:bg}, \triplet~localization area was observed eleven times in July 2020, except for shallower observations on July 6. Since the level 1 and 2 selections are utilized in each of the ZTF observations at multiple epochs, the TPR improves as the number of observations increases. On the other hand, the number of background objects also increases. Therefore, the epoch selection has been optimized by examining the manner in which the incorporation of multiple observations enhances the TPR in compensation of the background accumulation, using the ME1 and ME2 control samples (see also Section~\ref{sec:ana:multi}). For this multiple-epoch analysis, we use data from sky regions that were repeatedly observed over 11 epochs (from July 14 to 31, 2020) for the ME1 sample and 6 epochs (from July 14 to 23, 2020) for the ME2 sample (Figure~\ref{fig:overlap}). Figure~\ref{fig:tpr-nbg} shows the relation between the TPR and the number of background objects for different number of observations. Based on analysis using ME1 sample, we chose a set of 6 epochs from July 14 to July 23 that achieves a high TPR and a low number of background objects. Then, we analyze ME2 sample to improve the statistics in estimating the number of background objects.

\bibliography{sample7}{}
\bibliographystyle{aasjournalv7}



\end{document}